\documentclass[aps,prd,notitlepage,twocolumn,superscriptaddress,nofootinbib,10pt]{revtex4-1}
\usepackage[utf8]{inputenc}
\usepackage[english]{babel}
\usepackage{amsmath}
\usepackage{amsfonts}
\usepackage{amssymb}
\usepackage{listings}
\usepackage{lipsum}
\usepackage{multirow}
\usepackage{datetime}
\usepackage{graphicx}
\usepackage{mathtools}
\usepackage{mathrsfs}
\usepackage{dcolumn}
\usepackage{multirow}

\usepackage{color}
\usepackage{xcolor}
\usepackage{hyperref}
\hypersetup{
    colorlinks,
    linkcolor={red!50!black},
    citecolor={blue!50!black},
    urlcolor={blue!80!black}
}

\def\Lie{\mathcal{L}}
\def\p{\partial}

\begin{document}

\hfill {{\footnotesize USTC-ICTS/PCFT-24-24}}

\title{Complex structures of boson stars and anisotropic distribution of satellite galaxies}
\author{V\'ictor Jaramillo}
\email[]{jaramillo@ustc.edu.cn}
\affiliation{Department of Astronomy, University of Science and Technology of China, Hefei, Anhui 230026, China}
\affiliation{Department of Modern Physics, University of Science and Technology of China, Hefei 230026, China}

\author{Shuang-Yong Zhou}
\email[]{zhoushy@ustc.edu.cn}
\affiliation{Interdisciplinary Center for Theoretical Study, University of Science and Technology of China, Hefei, Anhui 230026, China}
\affiliation{Department of Modern Physics, University of Science and Technology of China, Hefei 230026, China}
\affiliation{Peng Huanwu Center for Fundamental Theory, Hefei, Anhui 230026, China}

\date{\today}

\begin{abstract} 

We construct and explore the complex structures of boson stars, drawing inspiration from similar configurations of non-topological solitons in Minkowski space. These ``molecular states'' of boson stars have a multipolar structure and both positive and negative Noether charges within one star, and the opposite charges swap with time. Thanks to the gravitational attraction, they exist even in the case of a free scalar field. We also explore the effects of scalar self-interactions on these complex structures. We propose to use  galactic-scale charge-swapping boson stars as a potential solution to the problem of the observed anisotropic distribution of satellite galaxies.

\end{abstract}

\maketitle

\tableofcontents

\section{Introduction}

Boson stars are nowadays well known solutions that arise in the context of complex scalar fields coupled to Einstein gravity. The first such self-gravitating, spatially localized, stationary configurations were obtained more than 50 years ago \cite{Kaup:1968zz}. This particular nonlinear solution, which nowadays is known as a mini-boson star, arise within the simplest model of a minimally coupled free scalar field and when spherical symmetry is assumed. Self-interactions such as the quartic \cite{Colpi:1986ye} and sextic \cite{Friedberg:1986tq} terms in the scalar potential enrich the family of boson stars. The variety of boson stars that can be found in the literature goes beyond modifications to the scalar potential, see e.g. \cite{Jetzer:1989av} for a generalization including electrically charged scalar fields, \cite{Herdeiro:2014goa} for solutions with horizons, \cite{Brito:2015pxa} for solutions of a vectorial bosonic field and some others to which we will refer later. 
However, many of them (and in particular the self-interacting ones) share similar dynamical properties with the mini-boson stars (see \cite{Liebling:2012fv} for a review on this topic and \cite{Zhang:2024bjo} for an unified view in the nonrelativistic limit), the most important of which include the stability and existence of a formation mechanism \cite{Seidel:1990jh,Balakrishna:1997ej,Seidel:1993zk,Bezares:2017mzk,Sanchis-Gual:2021phr,Jaramillo:2020rsv,Sanchis-Gual:2021edp,Bezares:2018qwa,Croft:2022bxq,Atteneder:2023pge,Sanchis-Gual:2018oui,Helfer:2021brt, Palenzuela:2006wp,Gleiser:1988ih,Cunha:2017wao,Kleihaus:2011sx, Tamaki:2010zz, Li:2019mlk}. 
This has enabled boson stars to find applications in various astrophysical and cosmological scenarios \cite{Olivares:2018abq,Amaro-Seoane:2010pks,Bernal:2006ci,Grandclement:2014msa,Suarez:2013iw,Hui:2016ltb,Jaramillo:2022gcq,Choi:2019mva,Solis-Lopez:2019lvz,Annulli:2020lyc,Guzman:2006yc,Guzman:2009zz,Zeng:2021oez,Rosa:2022tfv,Rosa:2023qcv,Bernal:2024wif} 

Stationary boson stars beyond spherical symmetry have also been studied for quite a long time. Among them are the (stationary) spinning boson stars \cite{Yoshida:1997qf}, which have a toroidal topology and are unstable \cite{Sanchis-Gual:2019ljs} unless self-interactions are included \cite{DiGiovanni:2020ror,Siemonsen:2020hcg}. Another example is that of (static) dipolar boson stars \cite{Yoshida:1997nd}. These solutions are balanced by the attraction of gravity and the repulsion of the two blobs, which possess the same value and sign of the Noether charge but with a $\pi$ relative phase.
They have been revisited recently in \cite{Cunha:2022tvk,Herdeiro:2023roz} and again, appear to be stabilized only when self-interactions are included \cite{Ildefonso:2023qty}. Beyond these two cases, there are the chains of boson stars, constructed naturally with self-interactions \cite{Herdeiro:2021mol,Gervalle:2022fze} and interestingly also without them \cite{Sun:2022duv,Sun:2023ord}. Additionally, there can be static (in terms of the spacetime metric)
multipolar boson stars \cite{Herdeiro:2020kvf} which contain both  dipolar components and chain components in addition to some non-axisymmetric structure, with the overall morphology of the energy density akin to the orbitals of the hydrogen atom. As the dipolar boson stars and chains of stars, the general multipolar case has a nonzero total Noether charge and every multipolar component contributes to the total charge with densities of the same sign.

Furthermore, boson star solutions by coupling  multiple fields to gravity have also been discussed. On the one hand, stationary non-trivial solutions in the free-field case have been obtained in \cite{Choptuik:2019zji} where the scalar fields oscillate with different frequencies. Ref. \cite{Bernal:2009zy} further considered a multi-state superposition. Ref. \cite{Bezares:2018qwa} studied the head-on collision of two spherical stars composed of different scalar fields. On the other hand, stationary solutions beyond spherical symmetry have also been obtained. Such is the case of the $\ell$-boson star \cite{Alcubierre:2018ahf} which considers a particular superposition of fields which preserves the spherical symmetry of the stress energy tensor; see \cite{Sanchis-Gual:2021edp} for a generalization that relaxes the constraint to obtain a family of solutions containing the $\ell$-boson star as well as the spinning and dipolar stars. The Newtonian version of this configuration was presented in Ref. \cite{Guzman:2019gqc}, where the generalization goes one step further and considers more general/excited radial profiles with multiple nodes. The morphology of this last multi-state multipolar configuration has interesting cosmological applications \cite{Solis-Lopez:2019lvz,Bernal:2024wif}. 

Q-balls \cite{Friedberg:1976me, Coleman:1985ki} (see \cite{Lee:1991ax} for a review) are the Minkowski space cousins of boson stars. Recently, it has been observed that for the model where spherical Q-balls exist, there also exist a tower of complex, multipolar Q-balls \cite{Copeland:2014qra}. These ``molecular states'' of Q-balls, despite being metastable, can live for a long time \cite{Copeland:2014qra, Xie:2021glp, Xie:2023psz}, especially for the logarithmic potential \cite{Hou:2022jcd}. These complex Q-balls have the intriguing feature that within a ball the co-existing positive and negative charges swap with time. The charge-swapping Q-balls are attractor solutions that can form from quite generic initial conditions \cite{Xie:2021glp, Hou:2022jcd}, and in particular they may form as a consequence of the Affleck-Dine(-like) condensate fragmentation in the early universe \cite{Hou:2022jcd, Zhou:2015yfa}, thanks to parametric resonance. On a slightly different note, it was recently observed that Q-balls and boson stars can induce superradiant amplification for scattering waves even in the spherically symmetrical cases \cite{Saffin:2022tub, Cardoso:2023dtm, Gao:2023gof}.

In light of the similarity of Q-balls and boson stars, in this paper, we shall investigate the complex, charge-swapping structures of boson stars where both positive and negative charges co-exist within a star. We will explore the free scalar field case and the polynomial sextic and logarithmic scalar potentials, illustrating the key differences between potentials and uncovering situations where the gravitational interactions are essential for the existence of such configurations. We will construct complex boson stars by preparing the initial data with superpositions of spherical boson stars located close to each other, and letting the configuration evolve and relax to form multipolar boson stars. These charge-swapping multipolar boson stars are different from the static multipolar boson stars \cite{Herdeiro:2020kvf} mentioned above. First of all, the charge-swapping boson stars are dynamical but with a well-defined morphology. However, the essential difference between the two is that: while in the static multipolar configurations the gravitational interaction is counteracted by a phase difference between the scalar field components, in the charge-swapping configurations it is the difference in the sign of the charge that allows the formation of the tower of multipolar configurations.

We also find that, while it is possible to achieve a considerable compactness, those charge-swapping boson stars in the large-size, low-compactness limit are well suited for a candidate for galactic dark matter halos. In such a scenario, the morphology of the dipolar configuration influences the structures moving in the outer region of the halo, and as a result, the galactic satellites moving around a host galaxy travel in coherently oriented planes. We shall explore the possibility of explaining the anisotropic distribution of satellite galaxies \cite{Pawlowski:2021ipt} by a charge-swapping dipolar galactic halo. These charge-swapping structures can be formed as a result of collisions of two structures with opposite charges, or, as mentioned, they can be formed by fragmentation of an Affleck-Dine-like condensate in the early universe.  

This paper is organized as follows. In Section \ref{sec:theory} we introduce the model, some relevant quantities used in the analysis and the explicit equations to treat charge-swapping boson stars as a Cauchy problem. After this, we also derive spherically symmetric boson stars and construct constraint-satisfying initial conditions after some clarification of the units and physical scales for our solutions. We then present and analyze the complex structures of boson stars for the minimally coupled free scalar theory in Section \ref{sec:freefield} and for self-interacting fields in Section \ref{sec:sifield}. In Section \ref{sec:sat_galaxies}, we will discuss the possibility of using the dipolar charge-swapping configurations as a potential solution to the anisotropic distribution of satellite galaxies in the Milky Way. We conclude in Section \ref{sec:conclusions}.

{\it Conventions}: The metric signature is $(-,+,+,+)$ and we choose the units with $c=1$. Many dimensionful quantities are written with tildes, while the corresponding dimensionless quantities are without tildes.

\section{Theory and setup}\label{sec:theory}

We focus on a canonical U(1) symmetric scalar $\Phi$ minimally coupled to gravity with action:
\begin{equation}\label{eq:action_EKG}
\tilde{S} = \int \text{d}^4 \tilde{x}\sqrt{-g}\left(\frac{\tilde{R}}{16\pi G}+\mathcal{L}_\Phi\right)\, ,
\end{equation}
where the Newton's constant is $G=1/m_P^2$ and
\begin{align}
  &\mathcal{L}_\Phi = -g^{\mu\nu}\frac{\partial \Phi^\dagger}{\partial\tilde{x}^\mu}\frac{\partial \Phi}{\partial\tilde{x}^\nu}-V(|\Phi|) \,.
\end{align}
We will consider multiple choices for the potential $V(|\Phi|)$ in this paper. We first consider the free, massive scalar case in Section \ref{sec:freefield}
\begin{align}
\label{phi2V}
    V(|\Phi|)=\mu^2 |\Phi|^2 \, ,
\end{align}
and in Section \ref{sec:sifield} we add two different self-interactions to the scalar
\begin{align}
\label{sexticpotdef}
  V(|\Phi|)&=\mu^2 |\Phi|^2 - \lambda |\Phi|^4 + \tilde{g} |\Phi|^6\, ,
  \\
  \label{logpotdef}
  V(|\Phi|) &= \mu^2|\Phi|^2\left(1+K\ln\frac{\, |\Phi|^2}{\mathcal{M}^2}\right) \, .
\end{align}
The sextic potential can be considered as a leading truncation for the tree-level effective potential, and the high order terms are neglected as they are further suppressed as compared to the leading terms. The logarithmic interaction (\ref{logpotdef}) is typical of quantum corrected effective potentials. As we see in Section \ref{sec:logpot}, the logarithmic potential is very similar to the free scalar case.

In the following of this section, we will use the sextic potential (\ref{sexticpotdef}) as an example to set up the computational framework. (While the free field case is a special case of the sextic potential, we will comment on the differences for the logarithmic case.) It is useful to define the following dimensionless quantities 
\begin{equation}
x^\mu = \mu \tilde{x}^\mu,~~ \phi=\frac{\sqrt{\lambda}\Phi}{\mu},~~ \mathfrak{g}=\frac{\tilde{g}\mu^2}{\lambda^2} ,~~ R =\frac{ \tilde{R}}{\mu^2} \,,
\end{equation}
similar to those used in \cite{Xie:2021glp}. 
With these dimensionless quantities, the action (\ref{eq:action_EKG}) with the sextic potential can be written as 
\begin{equation}\label{eq:action}
S =\lambda \tilde{S}  = \int \text{d}^4 x\sqrt{-g}\left(\frac{R}{16\pi \alpha_G}+\mathcal{L}_\phi\right)\, ,
\end{equation}
where
\begin{align}
  &\mathcal{L}_\phi = -g^{\mu\nu} \partial_\mu \phi^\dagger\partial_\nu \phi -V(|\phi|)\, , \\
  \label{eq:lagrangian}
  &\text{with}\quad V(|\phi|)= |\phi|^2 - |\phi|^4 + \mathfrak{g} |\phi|^6\, .
\end{align}
and we have defined an effective (dimensionless) gravitational constant 
\begin{equation}
\label{aiGdef}
\alpha_G = \frac{G \mu^2}{\lambda} =\frac{\mu^2}{m_P^2\lambda} \,,
\end{equation}
which indicates the strength of the gravitational self-interaction in the presence of the self-interacting scalar field. That is, we can simply use action (\ref{eq:action}) to extract physics for the sextic potential with generic couplings $\mu,\lambda,g$, different choices of $\mu$ and $\lambda$ corresponding to scalings of the solution obtained with action (\ref{eq:action}). Of course, for the classical solution to be valid, $\lambda$ should be relatively small, as its smallness can be linked to that of the reduced Planck constant: $\tilde{S}=S/(\lambda\hbar)$.

The free field potential (\ref{phi2V}) can be obtained from the sextic case by taking the limit $\alpha_G\to\infty$ (and $\tilde{g}\to 0$), where the $|\phi|^4$ and $\mathfrak{g}|\phi|^6$
terms become negligible, and the spherical solutions of the action \eqref{eq:action}
approach the mini-boson star sequence \cite{Kaup:1968zz,Ruffini:1969qy}.
In the opposite limit, when $\alpha_G\to 0$, the gravitational backreaction vanishes, the spacetime reduces to Minkowski space and the stationary solutions become the Q-balls.

For the logarithmic potential, the corresponding quantities are defined as follows
\begin{equation}
    x^\mu = \mu \tilde{x}^\mu,~~\phi=\frac{\Phi}{\mathcal{M}},~~ R =\frac{ \tilde{R}}{\mu^2} \,,
\end{equation}

Variation of the action \eqref{eq:action} with respect to $g_{\mu\nu}$ leads to the Einstein equations:
\begin{subequations}\label{eq:einstein}
\begin{eqnarray}
&& R_{\mu\nu}-\frac{1}{2} g_{\mu\nu} R= 8\pi\alpha_G T_{\mu\nu} ,
\label{eq:einstein2} \\
&& T_{\mu\nu}=g_{\mu\nu}\mathcal{L}_\phi-2\frac{\partial \mathcal{L}_\phi}{\partial g^{\mu\nu}},
\label{eq:einstein3}
\end{eqnarray}
\end{subequations}
and the $\phi$ variation leads to the Klein-Gordon equation,
\begin{equation}\label{eq:kg}
\nabla_\mu\nabla^\mu \phi = \frac{dV}{d(|\phi|^2)}\phi \, .
\end{equation}

Now, we define some global quantities that will be useful in both
the initial data construction and in the evolution of the dynamical system. The first one is the mass
of a stationary spacetime. While in Minkowski space one can simply use the integral of the energy density $T^{00}$
\begin{equation}\label{eq:E}
  E = \int_{\Sigma_t}T^{00}\ \sqrt{\gamma}d^3 x \, .
\end{equation}
as a measure of the total energy of the system, in curved spacetime other definitions need to be used to also account for the gravitational energy.
We shall use the Komar integral associated to the Killing vector $\xi = \partial_t$,
\begin{equation}\label{eq:komar}
M_{\rm Komar} = \frac{1}{4\pi\alpha_G} \int_{\Sigma_t} R_{\mu\nu} n^\mu \xi^\nu \sqrt{\gamma} d^3 x \, ,
\end{equation}
where $\Sigma_t$ is a spacelike surface of constant $t$ with $n^\mu$ a vector normal to it and
$\gamma$ is the determinant of the spatial metric induced in $\Sigma_t$.
We will further evaluate this integral in Section~\ref{sec:3p1}
when we discuss the 3+1 decomposition of the equations of motion. The Komar mass reduces to $E$ in Minkowski space for a stationary configuration. To see this, we use the Einstein equations, which gives
\begin{equation}\label{eq:komar2}
M_{\rm Komar} = 2 \int_{\Sigma_t} \left( T_{\mu\nu} n^\mu \xi^\nu - \frac{1}{2} T n_\mu\xi^\mu\right)  \sqrt{\gamma} d^3 x \, .
\end{equation}
Then we use the fact that the integral over the $T$ part of Eq.~(\ref{eq:komar2}) vanishes due to the
viral identity \cite{Herdeiro:2022ids}. Strictly speaking, $M_{\rm Komar}$ is not a conserved quantity in non-stationary spacetimes. However, it remains useful, along with $E$, for approximately tracking the evolution of a gravitational system (see, e.g., \cite{Croft:2022bxq, Jaramillo:2022zwg, Atteneder:2023pge}). Associated with the U(1) symmetry of the scalar field is the current
\begin{equation}
  j_\kappa = i\left(\phi^\dagger\nabla_\kappa\phi - \phi\nabla_\kappa\phi^\dagger\right) \, ,
\end{equation}
which is divergence-free and can be used to define a total scalar charge
\begin{equation}\label{eq:Q}
  Q = -\int n_\mu j^\mu\sqrt{\gamma}d^3 x \, .
\end{equation}

\subsection{3+1 decomposition}\label{sec:3p1}

As we wish to study the complex, dynamical structures of boson stars, we will need to evolve the Einstein-Klein-Gordon system.
This can be formulated as a Cauchy problem using a 3+1 decomposition
\begin{equation}\label{eq:3+1}
  ds^2=-\alpha^2 dt^2 + \gamma_{ij}\left(dx^i+\beta^idt\right)\left(dx^j+\beta^jdt\right) \, .
\end{equation}
Apart from the ``coordinate''-like variables $\gamma_{ij}$, the ``velocity''-like variables are given by the extrinsic curvature,
defined by,
\begin{equation}
  K_{ij}  =   - \frac{1}{2\alpha} \left( \partial_{t} - \Lie_{\beta} \right) \gamma_{ij} \, ,
\end{equation}
where $\Lie$ denotes the Lie derivative.
For the scalar field, we adopt the ``velocity''-like variable
\begin{equation}
\label{eq:Kphi}
K_{\phi} = -\frac{1}{2\alpha}  \left( \partial_{t} - \Lie_{\beta} \right) \phi \, .
\end{equation}

For a stable evolution of the gravitational system, we use the
Baumgarte-Shapiro-Shibata-Nakamura-Oohara-Kojima (BSSNOK)
formulation~\cite{Nakamura:1987zz,Shibata:1995we,Baumgarte:1998te}, which introduces auxiliary variables and  
makes use of a conformal metric with unit determinant 
${\tilde{\gamma}}_{ij}=\chi\,\gamma_{ij}$. In this approach, one re-writes the equations of motion as follows (see, e.g., \cite{alcubierre2008introduction} for details), 
\begin{subequations}
\label{eq:BSSNfull}
\begin{eqnarray}
\left( \partial_t -  \mathcal{L}_\beta \right)& \tilde \gamma_{ij} & = 
        - 2 \alpha \tilde A_{ij}\, , \\
\left( \partial_t -  \mathcal{L}_\beta \right)& \chi  & = 
        \frac{2}{3} \alpha \chi K\, , \\
\left( \partial_t -  \mathcal{L}_\beta \right)& K & = 
\chi\tilde{\gamma}^{ij}D_j D_i\alpha \nonumber\\
        & &  \hspace{-1.0cm}+ \alpha\left(\tilde{A}_{ij}\tilde{A}^{ij} + \frac{1}{3} K^2\right) + 4 \pi \alpha_G\alpha (\rho + S)\, , \\
\left( \partial_t -  \mathcal{L}_\beta \right)& \tilde A_{ij} & = 
        [\dots] - 8 \pi \alpha_G \alpha \left(
          \chi S_{ij} - \frac{S}{3} \tilde \gamma_{ij}
        \right)\, , \\
\left( \partial_t -  \mathcal{L}_\beta \right)& \tilde \Gamma^i & = 
        [\dots] - 16 \pi \alpha_G \alpha \chi^{-1} P^i\, ,\\
  \left(\p_t - \Lie_{\beta} \right)& \phi & = - 2 \alpha K_\phi \,, \\
  \left(\p_t - \Lie_{\beta} \right)& K_\phi &  = \alpha \left[ K K_{\phi} - \frac{1}{2} \chi\tilde{\gamma}^{ij} \tilde{D}_i \partial_j \phi 
  + \frac{1}{4} \tilde{\gamma}^{ij} \partial_i \phi \partial_j\chi\right. \nonumber \\
                & & \hspace{-1.5cm}  \left.+ \frac{1}{2}\left( 1 - 2 |\phi|^2 + 3\mathfrak{g}|\phi|^4 \right)\phi \right]
                 - \frac{1}{2} \chi\tilde{\gamma}^{ij} \partial_i \alpha \partial_j \phi
                 \, ,
\end{eqnarray}
\end{subequations}
where $K$ is the trace of $K_{ij}$, $\tilde{A}_{ij}$ is the traceless part of the conformal extrinsic curvature, $\tilde{D}_i$ (${D}_i$) is the covariant derivative compatible with $\tilde{\gamma}_{ij}(\gamma_{ij})$, $\tilde{\Gamma}^i$ are the conformal connection functions and we have defined the source terms
\begin{equation}
  \label{eq:source}
   \begin{aligned}
  \rho & \equiv T^{\mu \nu}n_{\mu}n_{\nu} \,,~~~~
  P_i  \equiv -\gamma_{i\mu} T^{\mu \nu}n_{\nu} \,, \\
  S_{ij} &\equiv \gamma^{\mu}{}_i \gamma^{\nu}{}_j T_{\mu \nu} \,, ~~~~
  S      \equiv \gamma^{ij}S_{ij} \,.
   \end{aligned}
\end{equation}
$[\dots]$ in the equation for $\tilde{A}_{ij}$ and $\tilde{\Gamma}^{i}$
denote the standard quantities defined in the right-hand side of the BSSNOK equations, except for the source terms
which have been explicitly included above. The real and imaginary parts of the scalar field $\phi$ are evolved separately in the actual code.
Finally, the evolution is subject to a set of constraints given by
\begin{align}
\label{eq:Hamiltonian}
H & \equiv {}^{(3)}R - K_{ij} K^{ij} + K^2 - 16 \pi\alpha_G\rho
       = 0\,,\\
\label{eq:momentumConstraint}
\mathcal{M}_{i} & \equiv D^{j} K_{ij} - D_{i} K 
        - 8\pi \alpha_G P_i
       = 0 \,.
\end{align}

\begin{figure*}[t!]
	\centering
		\includegraphics[width=0.443\textwidth]{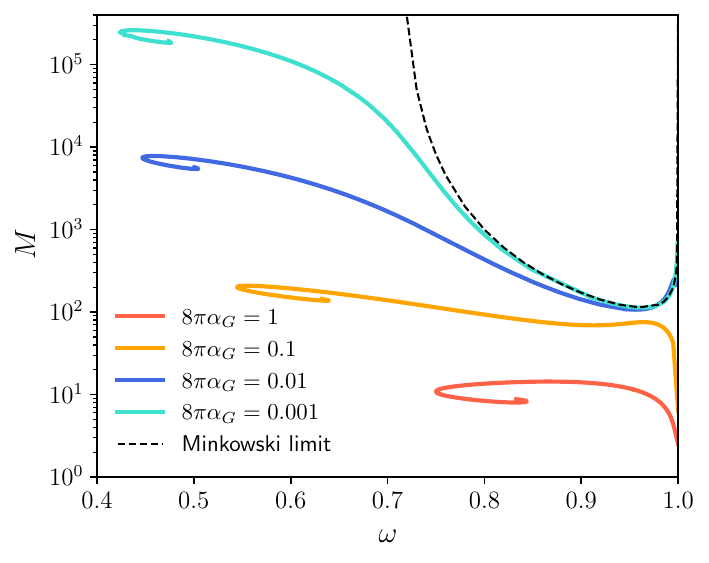} \quad \includegraphics[width=0.45\textwidth]{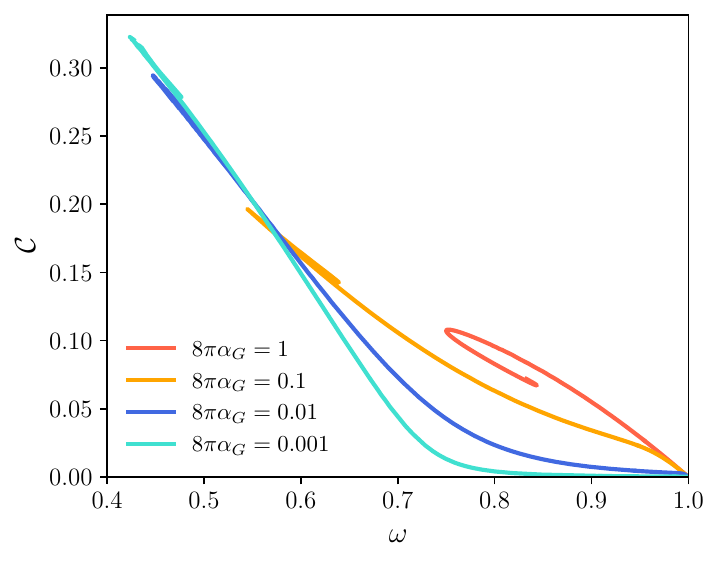}
	\caption{Dimensionless boson star mass $M$ (left) and compactness $\mathcal{C}$ (right) as a function of the frequency $\omega$
          for solutions with $\mathfrak{g} = 1/2$ and different values of $\alpha_G$. We see that $M$ converges to the Q-ball
          solution as $\alpha_G\to 0$. The extension of the stable branch for the cases with small
          (but not zero) $\alpha_G$ is notable. Also note that the compactness
          for these configurations can span (even within the stable branch)
          from very small values up to high values, such as 0.3.}
	\label{fig:Mass}
\end{figure*}

\subsection{Stationary isolated boson btars}

We first consider an isolated, spherically symmetric boson star solution
with a scalar field 
\begin{equation}
    \phi=f(r)\exp(-i\omega t) \,.
\end{equation}
We choose the following ansatz for the spherical and static line element,
\begin{equation}\label{eq:BS_metric}
ds^2=-\alpha^2(r)dt^2+\Psi^4(r)\left(dr^2+r^2d\Omega^2\right) \, .
\end{equation}
The solution for the functions $f$, $\alpha$, $\Psi$ and the eigenvalue $\omega$ can be obtained using
the spectral code described in \cite{Alcubierre:2021psa} to
solve the following equations
\begin{equation}\label{eq:BS_E}
\begin{split}
    &\Delta_3\Psi+2\pi\alpha_G\Psi^5\left[\left(\frac{\omega f}{\alpha}\right)^2+\frac{(\partial_r f)^2}{\Psi^4}+ V_f \right]=0,\\
    &\Delta_3 \alpha+\frac{2\partial_r \alpha\partial_r\Psi}{\Psi}-8\pi\alpha_G \alpha\Psi^4\left[2\left(\frac{\omega f}{\alpha}\right)^2 - V_f \right]=0, \\
    & \Delta_3 f + \frac{\partial_r f \partial_r \alpha}{\alpha}+2\frac{\partial_r f \partial_r \Psi}{\Psi}-\Psi^4\left[\frac{dV_f}{df^2}-\frac{\omega^2}{\alpha^2}\right] f=0 \ , 
\end{split}
\end{equation}
where $\Delta_3:=\partial_r^2 + \frac{2}{r}\partial_r $ and $V_f =  f^2 - f^4 + \mathfrak{g} f^6$. This system of ordinary differential equations should be supplied with the boundary conditions from the
regularity of the functions at $r=0$ and the asymptotic flatness of the spacetime:
\begin{eqnarray}\label{eq:bc}
    \partial_r f|_{r=0}= \partial_r \alpha|_{r=0}= \partial_r \Psi|_{r=0}=0\ ;\\
    f |_{r\to\infty} = 0,\quad \alpha|_{r\to\infty}= \Psi|_{r\to\infty}=1\ .
\end{eqnarray}

Now let us obtain the specific expressions of the global quantities that will be used to characterize the stars. For an isolated boson star, it is easy to extract the ADM mass $M_{\rm ADM}$, which
can be read off from the asymptotic behavior of the metric functions far away
from the center, $\Psi^4 = 1 + 2\alpha_G M_{\rm ADM}/r + \mathcal{O}(1/r^2)$. Alternatively, we
may evaluate the mass via the Komar integral \eqref{eq:komar}, which for the spacetime
\eqref{eq:BS_metric} reduces to
\begin{equation}
  M_{\rm Komar} = \frac{1}{\alpha_G}\lim_{r\to\infty}\partial_r (r^2\alpha) \, .
\end{equation}
This coincides with $M_{\rm ADM}$ given that the spacetime is stationary \cite{Gourgoulhon:2010ju}.
In fact, we will use the relative difference between the masses as an error indicator for the spherical solver \cite{Grandclement:2014msa}. We will use $M$ to refer to both masses in the following. For the U(1) charge of the scalar field, defined in Eq.~\eqref{eq:Q},
we get the following expression in the present case,
\begin{equation}
  Q = 2\int_{\Sigma_t}\frac{\omega f^2}{\alpha}\sqrt{\gamma}d^3 x \, .
\end{equation}
The radius $R_{99}$, commonly used to characterize the size of
a boson star, refers to the value of areal radius
that encompasses 99\% of the total boson star mass. More specifically, this radius is defined in coordinates \eqref{eq:BS_metric} as the value $R_{99}=\Psi^2(r_{99}) r_{99}$ such that the Misner-Sharp function $M(r)=-2r/(1+r\partial_r\ln\Psi)$ evaluated at $r = r_{99}$ is equal to $0.99M$. Finally, the compactness $\mathcal{C}$ of a boson star will be defined as the mass $M$ divided by $R_{99}$
\begin{equation}
    \mathcal{C} = \frac{M}{R_{99}} .
\end{equation}

For every pair of $(\alpha_G, \mathfrak{g})$, there exists a family of solutions parametrized
(in most cases) by the value of the scalar field at the center of the star, $f_0$.
For $\alpha_G\ll1$, we expect to recover the Q-ball solutions, and, in particular, the parameter space explored in \cite{Xie:2021glp}.
In Fig.~\ref{fig:Mass}, we plot the mass and the compactness for sequences of solutions with different values of coupling constant $\alpha_G$. From the left panel of Fig.~\ref{fig:Mass} we see that,
contrary to what happens in the Q-ball case,
the mass of the solutions is regularized at the maximal frequency $\omega$.
Additionally, when gravity is introduced, solutions beyond the Q-ball thin-wall limit can exist. In our case, this limit is $\omega = (1-1/(4\mathfrak{g}))^{1/2}$.
These solutions are of particular interest since, as can be seen in the right panel of Fig.~\ref{fig:Mass},
the compactness can reach very high values. For the corresponding static black hole, we have $\mathcal{C}=0.5$, while the maximum of $\mathcal{C}$ is around $0.1$ for mini-boson stars ($\lambda = 0 = \tilde{g}$) \cite{Jaramillo:2023twi}
and around $0.16$ for strongly self-interacting boson stars (large $\lambda/(G\mu^2)$, $\tilde{g}=0$) \cite{Amaro-Seoane:2010pks, Colpi:1986ye}. 

\begin{figure}[t!]
	\begin{centering}
		\includegraphics[width=0.45\textwidth]{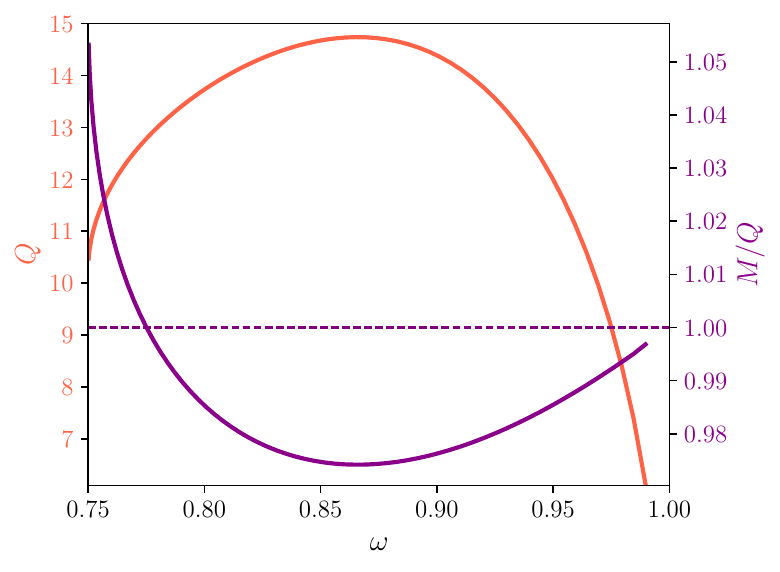} \includegraphics[width=0.455\textwidth]{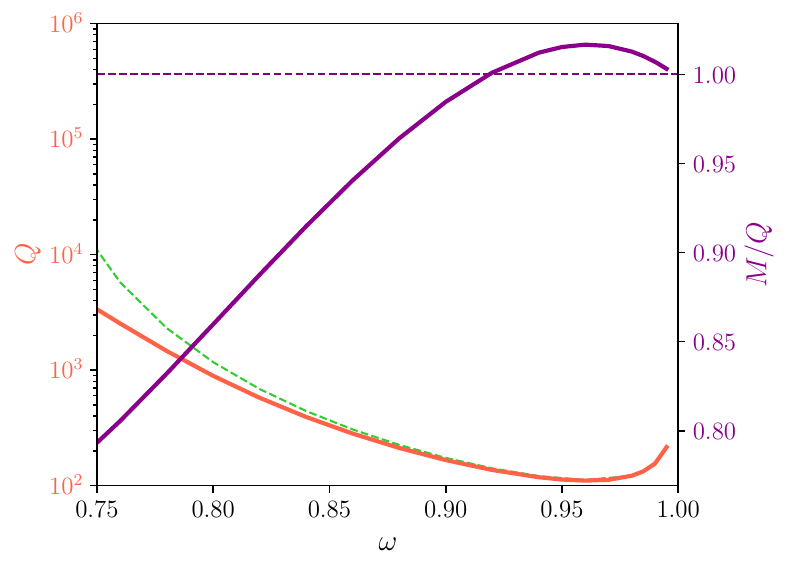}
	\end{centering}
	\caption{Sequence of isolated Q-stars with strong gravitational coupling $8\pi\alpha_G=1$ (top plot) and weak coupling $8\pi\alpha_G = 0.002$ (bottom plot). The value of the Noether charge for the corresponding $\alpha_G = 0$ (Q-ball) solution is also included as a green dotted line.}
	\label{fig:QandE}
\end{figure}

In Fig.~\ref{fig:QandE} we show a plot of the charge and energy-charge ratio for two
families with $\mathfrak{g} = 1/2$ and gravitational couplings $\alpha_G = 1/(8\pi)$ and $1/(8\pi\times 500)$.
From Fig.~\ref{fig:Mass} we already see that the effect of gravity substantially modifies the
configuration. In particular, when gravitational effects are strong (the top plot), the mass-charge ratio $M/Q$ goes above 1 when $\omega$ is close to the lower frequency limit. Note that $M/Q\leq1$ would be the stability condition of the Q-ball in the flat space limit, and for the Q-ball case, $M/Q\leq1$ is satisfied in the lower frequency limit, similar to the case of the bottom plot of Fig.~\ref{fig:QandE}. However, $M/Q\leq1$ may cease to be a reliable stability criterion in the presence of strong gravitational attractions.
Nevertheless, the solution with a
small value for $\alpha_G$ approaches the Q-ball solution as expected,
particularly in the region where the scalar field is smaller
and hence the gravitational backreaction is weaker. 

As mentioned in Section~\ref{sec:theory}, in the limit $\alpha_G\to\infty$, the model tends
formally to the case of a free scalar field, which gives rise to mini-boson stars with the scalar field
scaling inversely proportional to $\sqrt{\alpha_G}$ when $\alpha_G\gg1$. This tendency is consistent to what we see
as we increase the value of $\alpha_G$ in Fig.~\ref{fig:Mass} and also
Fig.~\ref{fig:numerical_application} in the next section.

\subsection{Units and scales}\label{sec:units}

\begin{figure*}
    \centering
		\includegraphics[width=0.45\textwidth]{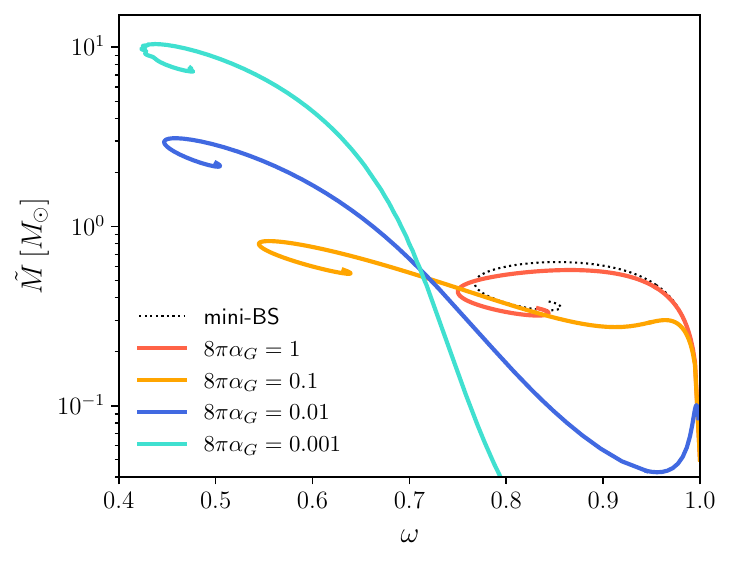} \quad\includegraphics[width=0.445\textwidth]{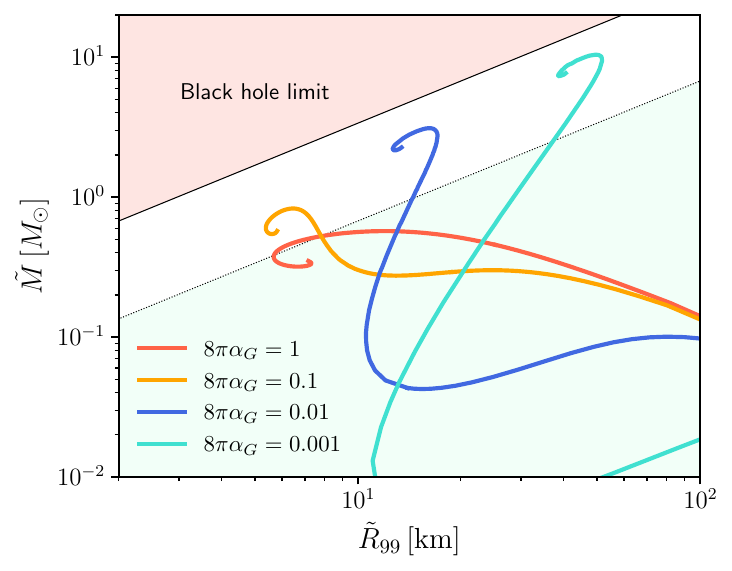}
	\caption{Sequence of solutions with $\mathfrak{g} = 1/2$ and for scalar mass $\hbar \mu = 1.33\times 10^{-10} \mathrm{eV}$.
          In the right plot, we indicate with a red band the region with compactness above the black hole limit $\mathcal{C}=1/2$. The big light blue band corresponds to the compactness bound of
          mini-boson stars.
          The models with
          smaller values of $\alpha_G$ 
          can reach higher values of compactness.
          The masses of most of the stable configurations in this example are of the order of the
          solar mass, while their radii are as small as tens of kilometers.
        For reference, the radius of the Sun is $R_\odot = 7\times10^{5}\mathrm{km}$.
        We can also see from both panels of this figure that the solutions
        approach the mini-boson star case as $\alpha_G$ increases.}
	\label{fig:numerical_application}
\end{figure*}

In the above formulation, the mass $M = M_{\rm ADM} = M_{\rm Komar}$ is dimensionless, so is the left part of the Einstein equations. To convert to the standard units, we can compare, for example, the asymptotic behavior
of the metric in different units, from which we find $\alpha_G M /r = G \tilde{M} / \tilde{r}$, with $\tilde{r}$ being
the dimensionful radius. Therefore, we can recover the standard units for the mass
\begin{equation}\label{eq:rerescaledM}
  \tilde{M} = \frac{\mu}{\lambda} M = \frac{\alpha_G}{G\mu} M \, .
\end{equation}
In units of the solar mass, we have 
\begin{equation}
\tilde{M} = \left(1.33\times10^{-10}\mathrm{eV}/(\hbar\mu) \right) \alpha_G M M_\odot \, .
\end{equation}
As one may expect, the Compton wavelength of the scalar $\mu^{-1}$ more or less sets the typical length scale of the localized configurations. Given the scalar mass $(\hbar\mu)$, say, in electronvolts, then the length scale in kilometers or in parsecs is respectively given by
\begin{align}
\mu^{-1}& = \left(1.97\times10^{-10}\frac{\mathrm{eV}}{\hbar\mu} \right) \mathrm{km}
\\
&= \left(6.40\times10^{-24}\frac{\mathrm{eV}}{\hbar\mu} \right) \mathrm{pc} \, .
\end{align}
These numerical factors can be used quickly to convert to the standard units for any distance quantity $\mu r$.
Both of them will be used in the following, km for astrophysical applications
and pc for the very dilute dark matter halo applications. See Fig.~\ref{fig:numerical_application} for an example of this conversion.

Smaller values of $\mu$ lead to configurations of the size of galactic dark matter halos.
For example, this is the case for the ultralight cosmological scalar field
\cite{Suarez:2013iw, Hui:2016ltb} with $\hbar\mu \sim 10^{-22}\mathrm{eV}$, implying a Compton wavelength of $\sim0.1$ pc and sizes for the corresponding cold and dilute self-gravitating objects of $\sim1$ kpc \cite{Hui:2016ltb}. 
We can also infer from Fig.~\ref{fig:numerical_application} that in such cases,
where an extremely low value of the compactness is expected, the present model with a variety of $\alpha_G$'s
can give rise to solitonic solutions that fulfill that requirement.

From $\alpha_G$'s definition \eqref{aiGdef}, we see that if the scalar coupling $\hbar\lambda$ is around $\mathcal{O}(0.1)$, then $\alpha_G$ is an extremely small number if $\mu^{-1}$ is of the kilometer scale, since $m_P\sim 10^{19}\,{\rm GeV}$. These cases essentially reduce to Q-balls, which for a kilometer scale $\mu^{-1}$ is astronomical in size. A larger $\alpha_G$ is obtained if either $\mu$ is close to the Planck mass or $\lambda$ is highly suppressed, say, around $\hbar\lambda\sim \mathcal{O}(10^{-10})$. For the former case, since $\mu$ sets the size of the localized configuration, these objects are quite small. The later corresponds to astronomical structures such as stars or larger objects.

\subsection{Preparing complex boson stars: solving initial constraints}\label{sec:constraint}

General Relativity is highly nonlinear, and the gravitational field in general does not obey a linear superposition law.
Thus, we can not simply superpose boson star solutions, which is in general not a solution, and let it relax toward
a complex boson star configuration. 
We need to carefully prepare the initial data by solving the Hamiltonian and momentum constraints.
The 3+1 formalism with a scalar field as a source was presented in Section~\ref{sec:3p1}.
For our purposes, we shall
choose $K_{ij}=0$ at $t=0$, which means that the momentum density $P^i$ of the initial
data that we intend to construct vanishes, that is, we wish to start with two clumps standing initially with
zero velocity. The extrinsic curvature
vanishes for time symmetric data, and therefore we can ignore the momentum constraints \eqref{eq:momentumConstraint} and solve only the
Hamiltonian constraint \eqref{eq:Hamiltonian}, which in this case is reduced to
\begin{align}
\label{eq:special_Hamiltonian}
H & \equiv {}^{(3)}\!R - 16 \pi \alpha_G \rho
       = 0\, .
\end{align}

We shall solve the Hamiltonian constraint by an iterative method. To that end, we start from the scalar configuration $f^{\rm (BS)}$ of a spherical boson star. With the $f^{\rm (BS)}$ configuration, as an initial guess, we
can prepare, for example, the dipolar scalar configuration by placing
a boson star and an anti-boson star on the $z$-axis separated by distance $d$,
\begin{equation}\label{eq:field_superposition}
  \phi = f^{\rm (BS)}(x,y,z-d/2)e^{-i\omega t}
                   + f^{\rm (BS)}(x,y,z+d/2) e^{i\omega t}.
\end{equation}
Similarly, for the canonical momentum of the scalar field, we use the initial input
\begin{equation}\label{eq:Kfield_superposition}
  K_\phi = K_\phi^{\rm (BS)}(x,y,z-d/2)
                    + K_\phi^{\rm (aBS)}(x,y,z+d/2).
\end{equation}
where $K_\phi^{\rm (BS)} = i\omega f^{\rm (BS)}e^{-i\omega t}/(2\alpha^{\rm (BS)})$ and $K_\phi^{\rm (aBS)} = - i\omega f^{\rm (BS)}e^{i\omega t}/(2\alpha^{\rm (BS)})$.
Then, we make a simple superposition of the corresponding gravitational fields. This setup gives us the initial guess in the Newton iterative scheme
used to solve the Hamiltonian constraint.
In the following we explain the procedure we follow to choose the initial guess for the spacetime metric and then obtain a solution to the Hamiltonian constraint.

We shall choose to solve Eq.~\eqref{eq:special_Hamiltonian}
using the conformal thin-sandwich approach, where the 3+1 spatial metric is related to
the conformal metric $\bar{\gamma}_{ij}$
by $\gamma_{ij}=\psi^4 \bar{\gamma}_{ij}$. Now, according to Eq.~\eqref{eq:BS_metric}, the spatial metric of an isolated boson
star in these coordinates is simply $\gamma_{ij}^{(\mathrm{BS})} = \Psi_{(\mathrm{BS})}^4\delta_{ij}$. Recall that the background metrics for an isolated boson star and anti-boson star are the same, so we choose the conformal metric $\bar{\gamma}_{ij}$ to be a simple superposition of the spatial metric \cite{Helfer:2021brt, Atteneder:2023pge},
\begin{equation}\label{eq:3metric}
\begin{split}
  \bar{\gamma}_{ij}(\mathbf{r}) &= A(\mathbf{r})^2 \, \delta_{ij}
\end{split}
\end{equation}
where we define the function $A$ in spherical coordinates\footnote{
Although in the dipolar case both $\phi$ and $K_\phi$
depend only on $r$ and $\theta$, we here
also keep track of the $\varphi$ angle dependence in the geometric and scalar field
quantities so that the generalization to the multipolar case
is straightforward to obtain from the equations presented here. 
For the dipole solutions, both the simple superposition guess $\bar{\gamma}_{ij}$
and the expected solution $\gamma_{ij}$ for the Hamiltonian constraint are
axisymmetric, so we also have $A = A(r,\theta)$.
}
as
\begin{equation}\label{eq:superpositionA}
\begin{split}
  A(r,\theta,\varphi) = &\left(\Psi_{(\mathrm{BS})}^4(x,y,z-d/2)\right.\\
  &\left.+ \Psi_{(\mathrm{BS})}^4(x,y,z+d/2) - 1\right)^{1/2} \, .
\end{split}
\end{equation}

The full spatial line element in spherical coordinates reads
\begin{equation}\label{eq:axi-3metric}
  dl^2 = \psi^4(r,\theta,\varphi) A^2(r,\theta,\varphi) \left(dr^2 + r^2d\Omega^2 \right) \, .
\end{equation}
In the conformal thin-sandwich formulation, the Hamiltonian constraint becomes the Lichnerowicz equation
\begin{equation}\label{eq:Lichnerowicz}
\tilde{D}^2\psi - 1/8\psi{}{}^{(3)}\! \tilde{R} = -2\pi\psi^5\rho \, ,
\end{equation}
where $\tilde{D}$ is the covariant derivative associated with $\bar{\gamma}$ and ${}^{(3)}\! \tilde{R}$
is the 3-dimensional Ricci scalar of $\bar{\gamma}_{ij}$. For the initial guess for the conformal factor, we choose $\psi(r,\theta,\varphi) = 1$, because of the conformal metric we have chosen.
Furthermore, we take as initial data for the lapse function $\alpha$ a simple superposition 
  \cite{Palenzuela:2006wp,Bezares:2018qwa,Sanchis-Gual:2018oui,Jaramillo:2022zwg}
$\alpha(r,\theta) = \alpha^{(\mathrm{BS})}(x,y,z-d/2) + \alpha^{(\mathrm{BS})}(x,y,z+d/2) - 1$. See Refs.~\cite{Helfer:2021brt,Atteneder:2023pge,Siemonsen:2023age,Evstafyeva:2022bpr} for other possibilities.

In order to solve the Hamiltonian constraint we calculate
\begin{equation}\label{eq:R3}
  {}^{(3)} \tilde{R} = -\frac{2}{A^3}\left(2\Delta_3 A - \frac{\partial A \partial A}{A}\right) \, ,
\end{equation}
and
\begin{equation}\label{eq:D2}
  \tilde{D}^2\psi = \frac{1}{A^2}\Delta_3\psi + \frac{1}{A^3}\partial A \partial \psi \, ,
\end{equation}
where we have used the 3-dimensional Laplacian and the scalar product of the gradients
in spherical coordinates
\begin{align}
  \Delta_3 &= \partial^2_r + \frac{2\partial_r}{r} + \frac{\partial_\theta^2}{r^2} + \frac{\partial_\theta}{r^2\tan\theta} + \frac{\partial_\varphi^2}{r^2\sin^2\theta} \, , \\
  \partial u \partial v &= \partial_r u \partial_r v + \frac{\partial_\theta u \partial_\theta v}{r^2}  + \frac{\partial_\varphi u \partial_\varphi v}{r^2\sin^2\theta}\, .
\end{align}
For the dipolar configuration, when applied to the fields $A$
and $\psi$, the derivatives with respect to $\varphi$ do not contribute.
All that remains is to obtain the energy density, which can be calculated from Eq.~\eqref{eq:source} 
using the metric of the spatial slice, leading to
\begin{equation}\label{eq:rho_superposition}
  \rho = 4 K_\phi K_\phi^\dagger + \frac{\partial \phi \partial \phi^\dagger}{\psi^4 A^2} + |\phi|^2 - |\phi|^4 + \mathfrak{g}|\phi|^6 \, .
\end{equation}
It is important to note that $\rho$ depends on $\psi$, so it must be updated at each iteration in the root-finding procedure used by the elliptic solver for the Hamiltonian constraint. This approach differs from the one typically employed in the context of fluid stars, where energy and momentum are fixed quantities, allowing for straightforward reconstruction of fluid variables. In our case, if $\rho$ is fixed, the scalar field variables can only be reconstructed after solving the Hamiltonian, which involves finding the scalar field that corresponds to the given $\rho$ through a differential equation. This process is complex and can result in initial data that do not accurately represent the physical situation of interest. Specifically, in charge-swapping configurations, we fix the scalar field initial data \eqref{eq:field_superposition}, \eqref{eq:Kfield_superposition} and treat them as the essential quantities we can control to achieve the desired dynamical configurations. As a consequence, $\rho$ becomes a derived quantity, which can lead to issues of non-uniqueness in the solutions of Eq.~\eqref{eq:Lichnerowicz}.

A comprehensive discussion of this issue in the context of initial data construction for boson stars can be found in Ref.\cite{Siemonsen:2023age}. In our implementation, we do not encounter alternative solutions and conduct three tests to assess the robustness of the code. First, we observe that the solutions for isolated boson stars ($\psi=1$) act as local attractors in the solution space, given the method used to solve the Lichnerowicz equation \eqref{eq:Lichnerowicz}. Second, we find that as the stars are moved farther apart, the solution converges to the $\psi=1$ case as the distance $d$ increases. Finally, we experiment with different implementations using superpositions other than those in Eq.\eqref{eq:superpositionA}, yet the (physical) metric coefficient $\psi^4 A^2$ consistently converges to the same solution.

In the following we describe in more detail how we solve the Lichnerowicz equation \eqref{eq:Lichnerowicz} in practice. Combining Eqs.~\eqref{eq:R3}, \eqref{eq:D2} and \eqref{eq:rho_superposition},
we  arrive at the following partial differential equation:
\begin{equation}\label{eq:eq_for_kadath}
\begin{split}
    \Delta_3\psi&  + \frac{\partial A\partial\psi}{A} \\
    &+ \frac{\psi}{4}\left(\frac{2\Delta_3 A}{A}-\frac{\partial A\partial A}{A^2}\right) + 2\pi\alpha_G\psi^5A^2\rho = 0 \, , 
\end{split}
\end{equation}
which needs to be solved with the following boundary conditions:
\begin{equation}
  \psi_{r\to\infty} = 1 , \quad \partial_r \psi|_{r=0} = 0, \quad \partial_\theta \psi|_{\theta = 0,\pi} = 0 \, .
\end{equation}

We solve this elliptic partial differential equation using the
\texttt{Kadath} library \cite{Kadath,Grandclement:2009ju}.
This spectral solver has been utilized in various theoretical physics scenarios, particularly for constructing axisymmetric boson stars \cite{Grandclement:2014msa,Jaramillo:2022gcq}.
Besides the axisymmetric dipolar case, we will also be interested in solving Eq.~\eqref{eq:eq_for_kadath} without
axisymmetry for the quadrupole and octupole cases. To this end, we employ the
\textit{spherical space} facility implemented in the library, which uses spherical coordinates
and decomposes the space into spherical shells with the possibility of placing as the
last shell one with a compactified radial coordinate such that boundary conditions
at infinity can be imposed exactly. Full details of this space can be found Ref. \cite{Grandclement:2009ju} where the \texttt{Kadath} solver was presented.

In the spherical space of the library, the spectral basis functions used in the decomposition of
the fields are chosen such that regularity
at the origin is ensured and the parity of the fields
with respect to $\theta=\pi/2$ plane is respected. In essence, the spectral code will
deal with
the spectral coefficients $F_{ijk}$ of certain function $F$.
If the function is unknown, the code will
try to find the solution through a Newton-Raphson iterative scheme
once an initial guess for $F_{ijk}$ or equivalently $F$ is given.
For our case, the unknown function is the conformal factor.
Roughly speaking, the fields are decomposed
into Chebyshev polynomials in the radial domain. In the $\theta$ domain, sines and cosines
of definite parity are used, and for the $\varphi$ angle, a simple Fourier
decomposition is utilized.
The particular relation between $F$ and $F_{ijk}$ is,
\begin{equation}
  \begin{split}
    F(r,\theta,\varphi) = &\sum_{i = 0}^{N_r}\sum_{j=0}^{N_\theta} \left[ \sum_{k=0\text{, $k$ even}}^{N_\varphi} F_{ijk} \cos(m\varphi) \right. \\
      &  + \left. \sum_{k=1\text{, $k$ odd}}^{N_\varphi}F_{ijk} \sin(m\varphi) \right] \Theta_j(\theta) X_i(x)
  \end{split}
\end{equation}
with $m = \left \lfloor{k/2}\right \rfloor$ and $x$ is some function that maps $r$ to
the domain $[-1,1]$, which depends on the particular radial domain considered.
In the inner domain, the (basis) functions $X_i$ and $\Theta_j$ are defined
according to the oddness of the integer $m$ and the symmetry of the function
$F$ with respect to the $\theta=\pi/2$ plane.
For $F$ symmetric,
\begin{eqnarray*}
  &X_i = T_{2i}(x)\, , \Theta_j = \cos(2j\theta)       &\text{if $m$ is even, } \\
  &X_i = T_{2j+1}(x)\, , \Theta_j = \sin((2j+1)\theta)   &\text{if $m$ is odd}.
\end{eqnarray*}
For $F$ antisymmetric,
\begin{eqnarray*}
  &X_i = T_{2i+1}(x)\, , \Theta_j = \cos((2j+1)\theta)       &\text{if $m$ is even, } \\
  &X_i = T_{2j}(x)\, ,  \Theta_j = \sin((2j)\theta)            &\text{if $m$ is odd}.
\end{eqnarray*}
For the outer domain,
the basis functions are simply chosen as
$X_i(x) = T_i(x)$ independently of the properties of $m$ and $F$.

The concrete form of the decomposition of the odd and even functions
is then used to transfer the initial data from the coefficient space
(3D array provided by \texttt{Kadath}) to a 3D
\texttt{Einstein Toolkit} mesh. For instance, in the case of the dipole,
the function $\phi$ and the metric coefficients are symmetric functions
with respect to the equatorial plane, but $K_\phi$ is anti-symmetric.

\section{Free scalar}\label{sec:freefield}

\begin{figure*}
    \centering
    \hspace{0.9cm}\includegraphics[width=0.14\textwidth]{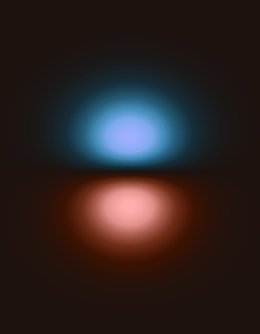}
    \includegraphics[width=0.14\textwidth]{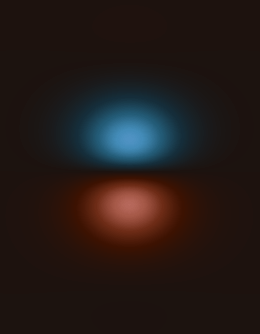}
    \includegraphics[width=0.14\textwidth]{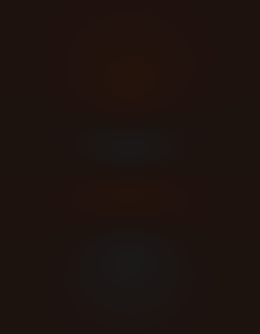}
    \includegraphics[width=0.14\textwidth]{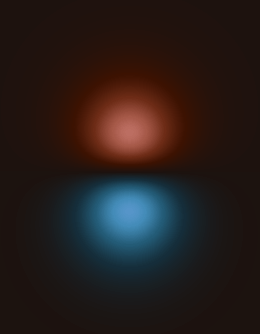}
    \includegraphics[width=0.14\textwidth]{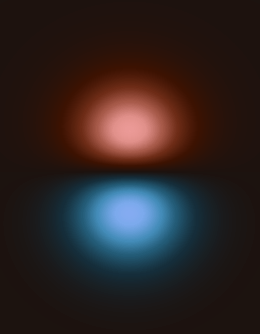}\\  
		\includegraphics[width=0.8\textwidth]{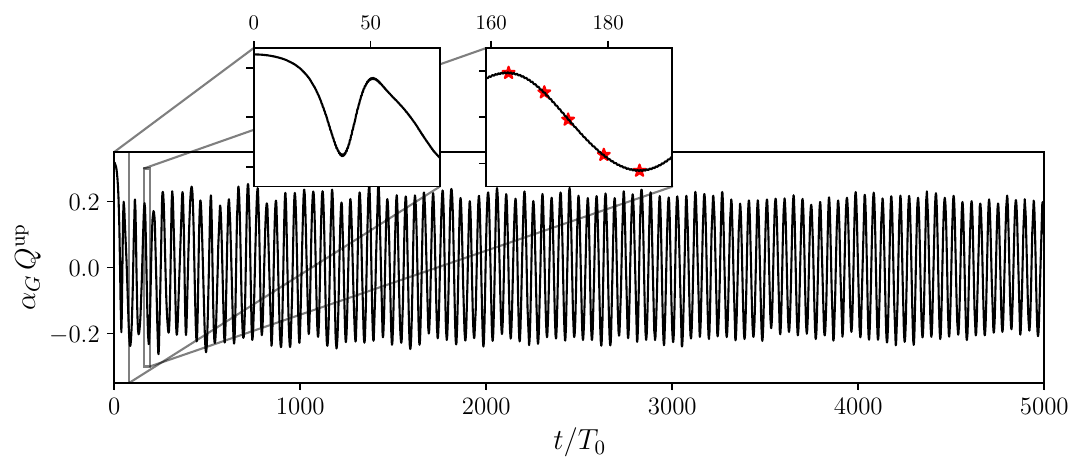}
  \caption{
    Evolution of charge density $-n_\mu j^\mu$ in a half charge-swapping period plotted in the
    $xz$ plane for a dipole configuration (top panel). The color scheme uses
    blue for positive charge, red for negative charge and black for zero charge, with
    the numerical values ranging $\pm3.6\times10^{-4}$.
    We also show the evolution of the total Noether charge in the $z>0$ half space, $Q^{\rm up}$,
    for the configuration initially constructed by superposing two boson stars with
    $\omega =\pm 0.98$ and separated by $d=24$ initially (bottom panel). The inset in the left shows the initial relaxation process. The points in the inset of the right
    show the five snapshots of the top panel. $T_0 = 2\pi$ is the characteristic time of
    the scalar field and also roughly the oscillation frequency of the two initial constituent stars. However, the charge-swapping frequency is much larger.
    The height of the boxes in the top panel snapshots is $\Delta z=28$.
    The inset can be seen in even greater detail in Fig.~\ref{fig:Q_FFT}. In accordance with footnote \ref{fn:limit} we use $\alpha_G$ to reconstruct the finite and physical quantity associated to the Noether charge.
  }
  \label{fig:Q_osc}
\end{figure*}

We shall first consider complex boson stars that are charge-swapping in the model where the scalar potential is simply
\begin{equation}
    V(|\phi|) =|\phi|^2 \,. 
\end{equation}
This can be formally obtained by taking the limit $\alpha_G\to \infty$ and $\mathfrak{g}/\alpha_G^2\to 0$ from the sextic potential.
The gravitational attraction makes it possible to have charge-swapping solitons even for this quadratic/free potential. In this model, stable (mini-)boson stars exist
from the zero mass limit, $\omega = 1$, up to the critical
mass \cite{Gleiser:1988ih} located at $\omega = 0.853$
and\footnote{\label{fn:limit}
In the limit $\alpha_G\to\infty$ the dimensionless mass $M$, energy $E$
and other quantities scale as $\alpha_G^{-1}$, so we
include $\alpha_G$ factors to make them
finite. However, it is important to say that
after restoring units with
the formulas given in Section~\ref{sec:theory},
the physical mass and Noether charge
of the solutions in the $\alpha_G\to\infty$ limit
correspond to the expected mini-boson star finite quantities (as they should). For
instance for the critical mass $\tilde{M} = 0.633/(G\mu)$,
as expected for the Kaup limit solution.} $\alpha_G M=0.633$; 
see Fig.~\ref{fig:numerical_application}. This stability range of boson stars
restricts the domain of configurations we can use to prepare
``molecular states'' of boson stars. 

At first glance, one might
expect that only by superposing configurations such that
the total mass does not significantly exceed the critical
value of an isolated boson star can such charge-swapping configurations be formed. However, on the initial relaxation stage, the system can radiate away a
significant amount of the initial energy, and therefore the charge-swapping
configuration can emerge from a large initial mass. 
Nevertheless, it is also true that if the initial mass exceeds a critical value, the final state typically is a black hole.
Considering this, we choose
six cases of mini-boson stars with frequencies between 0.95 and 0.995.
The information about these isolated boson stars is summarized
in Table~\ref{tab:ff_models}.
\begin{table}[b]
  \centering
  \begin{tabular}{l|ccccc|c}
    \hline\hline
    $\omega$ & $\alpha_GM$ & $\alpha_GQ$ & $R_{99}$ & $\mathcal{C}$ & $\alpha_GE$ & $\sim d_{\rm min}$ \\ \hline
    $0.95$   & $0.490$     & $0.497$     & $17.4$  & $0.0296$      & $0.388$     & $>36$\\
    $0.96$   & $0.450$     & $0.455$     & $18.9$  & $0.0235$      & $0.373$     & $36$\\
    $0.97$   & $0.399$     & $0.403$     & $21.7$  & $0.0186$      & $0.347$     & $24$\\
    $0.98$   & $0.334$     & $0.336$     & $27.3$  & $0.0123$      & $0.304$     & $18$\\
    \hline
    $0.99$   & $0.242$     & $0.243$     & $40.6$  & $0.0058$      & $0.231$     & -\\
    $0.995$  & $0.173$     & $0.174$     & $55.4$  & $0.0032$      & $0.171$     & -\\
    \hline\hline
  \end{tabular}
  \caption{Classical observables of the mini-boson stars used in the preparation of
     charge-swapping boson stars. All of them correspond
    to low-compactness stars. The value of $E$ is calculated using the integral
    \eqref{eq:E}, despite that it only characterizes the total energy of the
    solution in the weak-field limit. The value $d_{\rm min}$ in the last column
    corresponds to the minimal distance between the centers of the
    boson stars in the dipolar case such that a long-living boson-anti boson star configuration can form The configurations with ``-'' in the last column, represent cases not explored or discussed in the dipolar scenario.
    }
  \label{tab:ff_models}
\end{table}

To determine how the system evolves, the initial data are exported into the evolution scheme Eqs.~\eqref{eq:BSSNfull},
implemented in the open source \texttt{Einstein Toolkit}
infrastructure \cite{EinsteinToolkit:2023_11,Loffler:2011ay} using the
\texttt{McLachan} \cite{Brown:2008sb} thorn to evolve the metric quantities and a modified
version of the \texttt{Scalar} thorn \cite{Cunha:2017wao,Canuda:zenodo} to evolve
the scalar field. To track the formation of apparent horizons we use the
\texttt{AHFinderDirect} \cite{Thornburg:2003sf}. The evolution is implemented with a grid
spacing of $\Delta x^i = 1.6$ and a Courant factor of $\Delta t/\Delta x=0.125$ and with different
fixed refinement levels included in the \texttt{Carpet} thorn \cite{Schnetter:2003rb}.
In all cases, the size of the computational domain box is at least twice that of
$d + 2 R_{99}$ (the approximate size of the configuration at $t=0$)
for the configuration with the highest compactness and largest $d$.

The particular size of the domain and the distribution of the refinement level
will be specified in the following and kept fixed for all the simulations in the same
section for an easy comparison.
As discussed in \cite{Xie:2021glp} and \cite{Hou:2022jcd}, boundary conditions are
important when investigating the lifetime of a charge-swapping configuration.
In line with this, we have also employed absorbing (radiative) boundary conditions that are usual for boson star simulations. For the scalar field, we
use the standard Sommerfield boundary conditions
but additionally include the first order correction
to the massless scalar field dispersion relation \cite{Seidel:1990jh}:
$\partial_t K_\phi = -\partial_rK_\phi- K_\phi/r+\mu^2\alpha\phi/4$
at $r\to\infty$. As for the gauge freedoms, we impose the
``Gamma-driver'' condition \cite{Alcubierre:2002kk} for the shift vector $\beta^i$
and the ``1+log'' condition \cite{Bona:1994dr} for the lapse $\alpha$.

The interpolation from the initial data spectral coefficients and the analysis
of various quantities during the evolution are done using
separate thorns also within the \texttt{Cactus} framework. The choices of the numerical
setup employed in this work have been tested in the Appendix.

\subsection{Dipolar (charge-swapping) boson star}

\begin{figure*}
  \centering
    \includegraphics[width=0.44\textwidth]{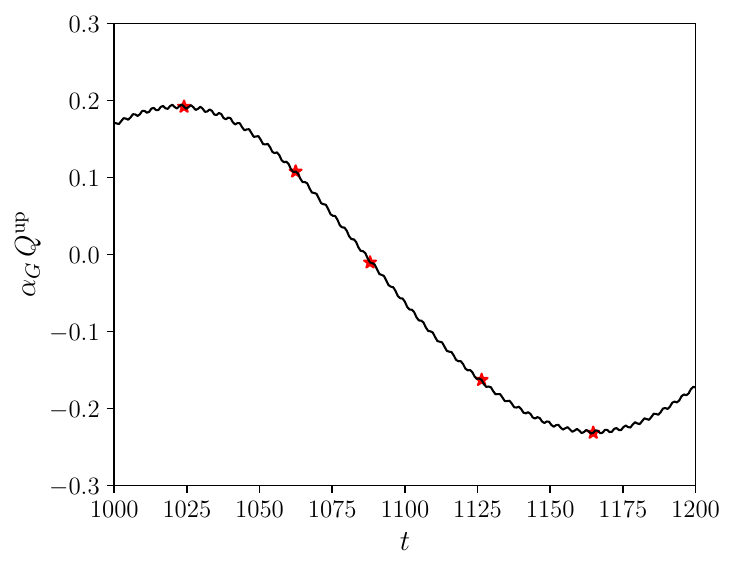}
    \includegraphics[width=0.425\textwidth]{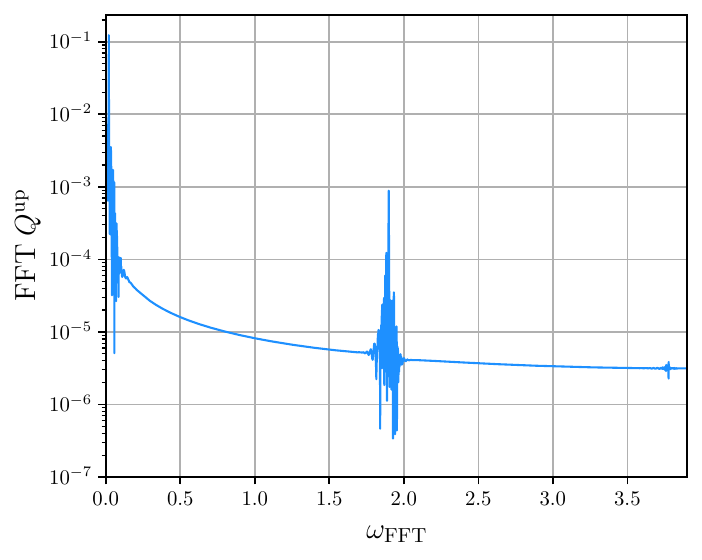}\\
    \includegraphics[width=0.43\textwidth]{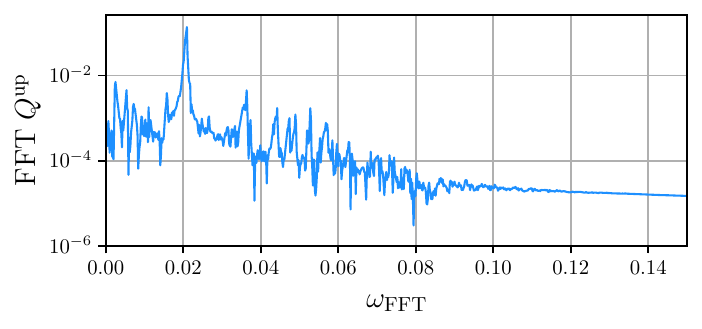}\hspace{0.3cm}
    \includegraphics[width=0.44\textwidth]{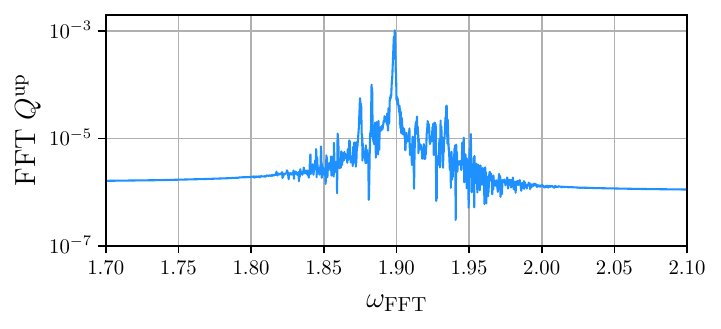}
  \caption{Amplification of the inset in Fig.~\ref{fig:Q_osc} (top left panel)
    and Fourier transform of $Q^{\rm up}$ for the temporal waveform in Fig.~\ref{fig:Q_osc}.
    The two main peak regions are amplified in the bottom panels.
  }
  \label{fig:Q_FFT}
\end{figure*}

For the dipolar boson star, we initially prepare a superposition of two spherical boson stars with opposite charges but
the same values of $|\omega|$, which take the first four
rows of Table~\ref{tab:ff_models}, corresponding to $\omega = 0.95$, 0.96, 0.97 and 0.98.
Then, following the procedure presented in Section~\ref{sec:constraint},
there is only one more parameter to choose: the initial distance between the superposed stars $d$.
We vary $d$ between 0 and 36 in steps of 6 for the four frequencies. 
For the low compactness case with $\omega=0.98$, which is our fiducial model, we also explore $d=42$.
This gives a total of 25 simulations for the free-field dipole case,
for which we choose a cubic grid of side 160 and use the mirror-symmetry
in the plane $xz$ and $yz$ to reduce the computational cost by a factor of 4.
We use three refinement levels with a grid-spacing of $\Delta x = 0.4$ in the finest grid, refining the grid in each level by a factor of two.

We find that if the initial distance is too small, or the solution too compact,
an apparent horizon is formed and the scalar field configuration results in 
gravitational collapse within $t\sim10^3$. For a fixed value of $\omega$,
we observe that as the distance $d$ increases, a greater amount of energy is radiated away in the initial relaxation period. Nevertheless, if the stars in Table \ref{tab:ff_models} are placed sufficiently far away from each other initially, they can attract and relax to form a charge-swapping configuration.

A typical scenario of a charge-swapping boson star forming
is shown in Fig.~\ref{fig:Q_osc}. The first observation is that, unlike their flat spacetime counterparts with a polynomial potential \cite{Xie:2021glp}, there is no indication of decay or a subsequent final oscillaton stage with total $Q = 0$. For all configurations where a charge-swapping boson star successfully forms in the free field case, the evolution exhibits only two qualitatively distinct stages: an initial relaxation phase, followed by a charge-swapping plateau.

The first stage is depicted in the left inset of Fig.~\ref{fig:Q_osc}. In this specific case, we consider initial boson stars with a radius of  $R_{99} = 27.3$ , initially separated by a distance of  $d = 24$ . Consistent with these parameters, the charge swap begins promptly after  $t = 0$, indicating that the initial configuration is relatively close to the charge-swapping bound state. We observe that the U(1) charge in the  $z > 0$  half-space,  $Q^\text{up}$ , starts oscillating with an amplitude slightly smaller than the initial one. This reduction is due to a small portion of the scalar field being ejected shortly after the simulation begins. Although, in this example, the stars are already very close to each other at  $t = 0$, it is important to say, as we will discuss later, that in some cases, positioning the stars farther apart is crucial for achieving a charge-swapping pattern. In these cases we will refer to the early part of the initial relaxation stage, when the stars meet and start interacting both gravitationally and through scalar field interactions leading to the interchange of Noether charge, as the merger phase.

In the top panel of Fig.~\ref{fig:Q_osc}, we show half of a charge-swapping period, with the first plot and the last plot showing that the positive and negative charges have been completely inverted during this half period.

For $\omega = 0.98$, we find that below the value $d=24$,
which is the one showed in Fig.~\ref{fig:Q_osc}, the next case,
which is $d=18$, also forms a regular horizonless final configuration, but
the configuration with $d=12$ forms a black hole. So according to the
mapping of the parameter space we use, the minimum distance to form a
boson-anti boson regular system is $d_{\min}=18$. For $\omega = 0.97$, which is
both more massive and compact, we find a bigger minimum distance
$d_{\min}=24$ is required. The trend continues with the $0.96$ and $0.95$
cases as displayed in the last column of Table \ref{tab:ff_models}.
Of course, the configurations with smaller $\omega$ are initially smaller and, as they require a larger initial distance to avoid collapse, the merger of the stars is delayed. For these cases, and before the merger, the charge $Q^{\mathrm up}$ remains constant, then decreases (at the merger) and starts to oscillate regularly once the initial relaxation period (which can be violent) is over. The process is similar to what we will show to happen in the quadrupolar case in Fig.~\ref{fig:Q_quad} below.

The charge-swapping configuration analyzed in Fig.~\ref{fig:Q_osc} and all the other
charge-swapping configurations we surveyed have very long lifetimes.
Given that $T_0=2\pi$ is approximately the oscillation period of the scalar field,
we find that the complex boson star survives for at least 5000 $T_0$.
For a selection of these simulations, we have let them evolve for
three times as long, and we find that they continue oscillating without decay. 

The charge distribution of the formed configurations contains
some strong Fourier modes which are visible immediately if we combine
Fig.~\ref{fig:Q_osc} together with the first panel in Fig.~\ref{fig:Q_FFT}.
As was explicitly demonstrated already in Ref. \cite{Xie:2021glp} for the Q-ball case,
the most important frequency is related to the global difference between the
real and the imaginary part of the scalar field. In the boson star case, this is
induced not by a nonlinear scalar potential, but by the gravitational effects.
In Fig.~\ref{fig:Q_FFT}, we also show the temporal Fourier transforms of
$Q^{\rm up}$ (top right panel), for the waveform presented
in Fig.~\ref{fig:Q_osc}. We also see from the bottom panels of Fig.~\ref{fig:Q_FFT}
that the period of charge swapping is around $T_{\rm swap} = 2\times 150$, since the dominant peak is around
$\omega_{\rm FFT}=0.02 \approx 2\pi/300$. The other low frequency modes,
which modulate the amplitude of $Q^{\rm up}$, are present to the
left of this dominant peak (see the bottom left panel of Fig.~\ref{fig:Q_FFT}).
Similar to the flat spacetime charge-swapping solutions
\cite{Copeland:2014qra, Xie:2021glp}, we see a high frequency oscillation
which corresponds to the small undulations visible in the top left panel of
Fig.~\ref{fig:Q_FFT}, having a peak frequency close to $\omega_{\rm FFT}=1.9$
(see the bottom right panel).

\begin{figure*}
  \begin{centering}
    \includegraphics[width=0.46\textwidth]{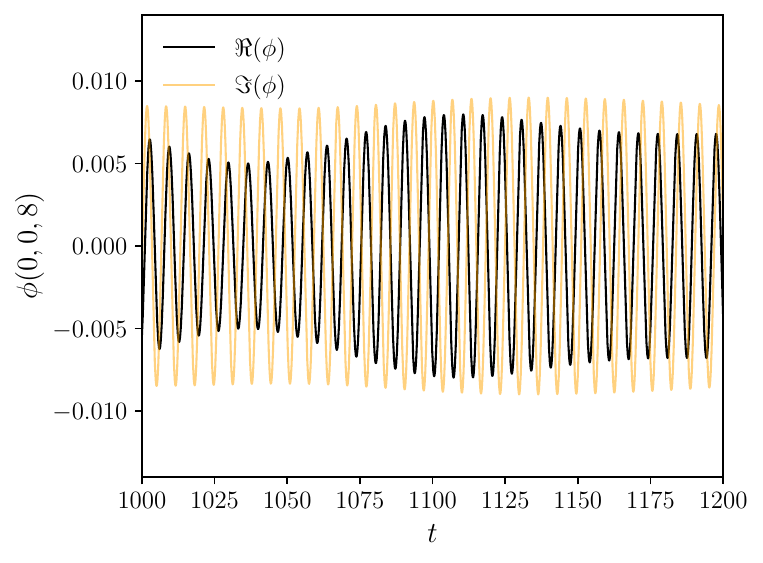}
    \includegraphics[width=0.44\textwidth]{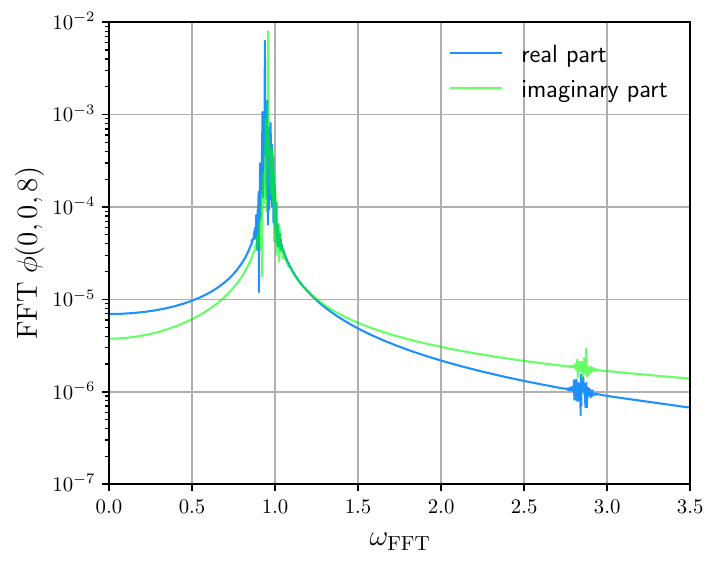}
  \end{centering}
  \caption{
    Real and imaginary part of the scalar field in the same time window as
    the half-period inset of Fig.~\ref{fig:Q_osc} and evaluated at $(x,y,z)=(0,0,8)$
    (left panel). The right panel is the Fourier decomposition of the scalar field, the blue line corresponding
    to the (fast) Fourier transform of the real part, while the green one to the imaginary part. The maxima
    are at $\omega_r = 0.939$ and $\omega_i=0.959$ respectively.
    The phase difference between
    $\Re(\phi)$ and $\Im(\phi)$ is changed in this time window
    between $t=1000$ and $t=1200$.
  }
  \label{fig:phi_FFT}
\end{figure*}

One can verify that the difference between the
frequencies of the real and imaginary component of the scalar field is the origin
of the dominant oscillation mode of the charge distribution.
To see this, we make a spectral decomposition of the scalar
field at a point outside the origin---at the origin, the imaginary
part of $\phi$ is zero by construction. We see from Fig.~\ref{fig:phi_FFT}
that the scalar field is practically monochromatic and the difference
between the maximum of the real and imaginary parts gives $\omega_R - \omega_I = 0.020$
(see the caption of Fig.~\ref{fig:phi_FFT} for more details), which is consistent
with the dominant peak of $Q^{\rm up}$. In addition, we also find a second
peak which is located around the first odd multiple of the base
frequency, analogously to what was found in \cite{Xie:2021glp} and consistent
with the fact that a scalar potential has a $\mathbb{Z}_2$ symmetry.
On the other hand, the small oscillations of $Q^{\rm up}$ that we already discussed
corresponding the peak in the bottom right panel of Fig.~\ref{fig:Q_FFT},
coincide very well with $\omega_R +\omega_I = 1.898$
and can be shown to be produced by the next
order perturbation term on the charge density $\propto j^0$, after
the term with frequency $\omega_R - \omega_I$.

The other free-field dipolar solutions also have a clear
harmonic dependence in time, so we have extracted the averaged period of
$\phi$ for these cases and displayed the information in Table \ref{tab:periods}.
\begin{table}[b]
  \centering
  \begin{tabular}{cc|cc}
    \hline\hline
    $\omega$ & $d$  & period  & afreq. \\ \hline
    $0.98$   & $18$ & $6.783$ & $0.926$  \\
    $0.98$   & $24$ & $6.697$ & $0.938$  \\
    $0.98$   & $30$ & $6.642$ & $0.946$  \\
    $0.98$   & $36$ & $6.609$ & $0.951$  \\
    $0.98$   & $42$ & $6.588$ & $0.954$  \\
    &&&\\
    $0.97$   & $24$ & $6.942$ & $0.905$  \\
    $0.97$   & $30$ & $6.858$ & $0.916$  \\
    $0.97$   & $36$ & $6.810$ & $0.932$  \\
    &&&\\
    $0.96$   & $36$ & $7.132$ & $0.881$  \\
    \hline\hline
  \end{tabular}
  \caption{Averaged period for $\Re\,\phi(0,0,0)$ for several free-field cases, considering various different initial separations.
    We only list the simulations where a charge-swapping configuration
    is formed. The period is defined as the time elapsed between two
    zero points of $\phi(0,0,0)$ where $\partial_t\phi(0,0,0)>0$, averaged over the
    data from $t=10^2 T_0$ to $t=5\times 10^3 T_0$. The angular frequency
    (afreq.) refers to $2\pi$ divided by this averaged period.
    }
  \label{tab:periods}
\end{table}
We see that for a given value of the (initial) frequency $\omega$,
the angular frequency of the scalar field in the formed configuration
increases with $d$ (more energy is radiated away during the
initial relaxation phase) and the configuration
evolves to a ``more Newtonian'' configuration. Also, notice that the
angular frequency is always smaller than the corresponding value at $t=0$, $\omega$.
This is actually opposite to the behavior of single unstable boson star
undergoing gravitational cooling \cite{Seidel:1993zk} and tending to a stable configuration.

\begin{figure}
	\begin{center}
		\includegraphics[width=0.45\textwidth]{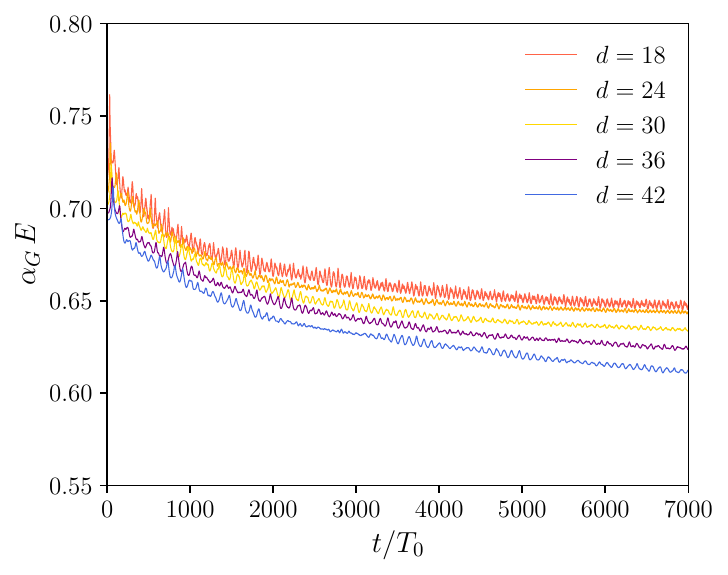}
	\end{center}
	\caption{
 Energy functional defined in Eq.~\eqref{eq:E} for several free scalar field dipoles using different initial distances for configurations prepared with $\omega = 0.98$.}
	\label{fig:results_Emini}
\end{figure}

In addition to tracking the evolution of $Q^{\rm up}$, we utilize the integral described in Eq.~\eqref{eq:E} to monitor the evolution of the systems. We find that 
the integral $E$, plotted in Fig.~\ref{fig:results_Emini}, converges gradually to a nonzero value (at least for the time probed by our simulations). Specifically, for a fixed $\omega$, $E$ decreases as $d$ increases.

\subsection{Higher multipolar boson stars}

Dipolar boson stars formed after a head-on collision of
two boson stars were previously observed and shown to survive at least until $t \sim 200\, T_0$ \cite{Palenzuela:2006wp}. In the previous subsection, we observed that these objects can actually exist for much longer periods and display an intriguing charge-swapping pattern. However, to the best of our knowledge,
no tower of composite stars with multipolar distribution has been reported.
In the case of Q-balls, these configurations do exist \cite{Copeland:2014qra}, and we now examine whether
it is possible to generalize the dipolar case to get configurations formed of positive
and negative charge pairs with more general morphologies.
To this end, in this subsection, we consider superpositions of four as well as eight
stars with equal and opposite charges, and place them in quadrupole and octupole configurations respectively.

In Section~\ref{sec:constraint}, we have used the dipolar case as an example to illustrate the preparation of the initial data for the simulations. In particular, we specified how to
solve the Hamiltonian constraint. For the quadrupolar case, the procedure is similar. Now, we can place spherical
stars centered at the vertices of a square with side length $d$ and positioned in the $xz$ plane:
\begin{equation}\label{eq:phi_quad}
  \begin{split}
    \phi &= f^{\rm (BS)}(x-d/2,y,z-d/2) e^{-i\omega t} \\
         &~~~ + f^{\rm (BS)}(x-d/2,y,z+d/2) e^{i\omega t} \\
         &~~~ + f^{\rm (BS)}(x+d/2,y,z-d/2) e^{i\omega t} \\
         &~~~ + f^{\rm (BS)}(x+d/2,y,z+d/2) e^{-i\omega t} \, .
  \end{split}
\end{equation}
Consequently, for the canonical momentum, we impose
\begin{equation}\label{eq:kphi_quad}
  \begin{split}
    K_\phi &=K_\phi^{\rm (BS)}(x-d/2,y,z-d/2) \\
           &~~~ + K_\phi^{\rm (aBS)}(x-d/2,y,z+d/2) \\
           &~~~ + K_\phi^{\rm (aBS)}(x+d/2,y,z-d/2) \\
           &~~~ + K_\phi^{\rm (BS)}(x+d/2,y,z+d/2) \, .
  \end{split}
\end{equation}
After this, the generalization for the superposition of the conformal
metric and the lapse function is straightforward.
The rest of the procedure is exactly the same,
since we have formulated the equations between Eq.~\eqref{eq:axi-3metric}
and Eq.~\eqref{eq:eq_for_kadath} without assuming
any particular symmetry. 

The orientation of the square defined by the centers of the stars at $t=0$, as given by Eq.~\eqref{eq:phi_quad}, is selected to minimize adjustments in the implementation within \texttt{Kadath}, particularly concerning the parity of the fields used in the solution. However, to make
better use of the Einstein Toolkit resources, we choose to perform a transformation
of the coordinates $x^i = \Lambda^i_{i'} {x'}^{i'}$ with $\Lambda$ being the rotation
matrix around the $y$-axis $R_y(\delta)$, $\delta$ being the rotation angle. Taking $\delta = \pi/4$,
the system has mirror-symmetry\footnote{
In the thorn arrangement used for the simulations,
only mirror-symmetry itself can be imposed on the scalar field, but not the anti-symmetry.
}
with respect to the three $x^i=0$ planes and
the computational cost is reduced by a factor of eight. So this is actually faster than the dipole case.
Nevertheless, the quadrupolar configurations are lager in size. Thus, for these simulations, we double the mesh size and the refinement level boundaries.

\begin{figure}
	\begin{center}
		\includegraphics[width=0.45\textwidth]{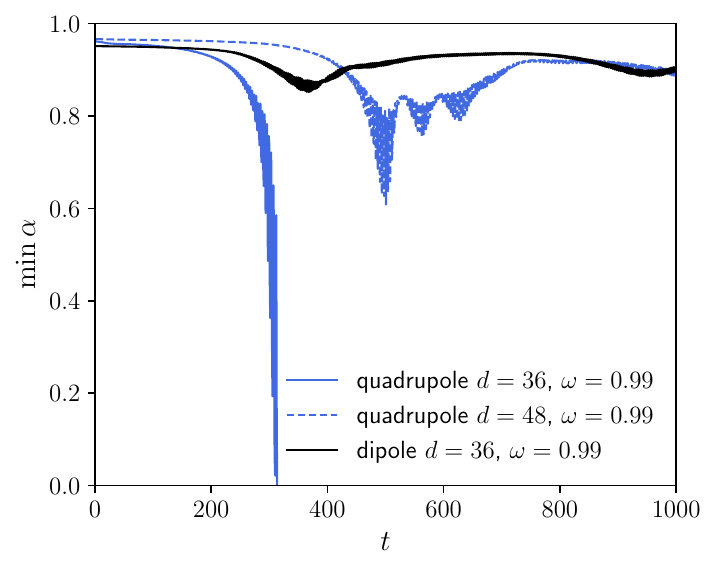}
	\end{center}
	\caption{Minimum value of the lapse function for two free field quadrupole cases
          with $\omega = 0.99$ and two different values of $d$. For comparison, we add the
          curve of the $\min\alpha$ for the dipolar case with the same values of $d$ and $\omega$
        as the collapsing configuration.}
	\label{fig:collapse_lapse}
\end{figure}

We find that none of the first five types of spherical stars in Table~\ref{tab:ff_models} leads to charge-swapping configurations for
the same range of $d$ as in the dipole case, not even extending up to a value of $d=48$. Black holes are often formed instead.
In contrast, the last
two types of stars with frequencies of $0.99$ and $0.995$ are capable of
forming non-trivial configurations that are everywhere regular. Stationary boson stars
with these frequencies are very dilute, as can be seen from their values of $\mathcal{C}$
and radius $R_{99}$ in Table~\ref{tab:ff_models}. We have tried three
$d$ distances for these two $\omega$'s respectively: 48, 42 and 36, and find that for both
$\omega$'s no apparent horizon is formed for $d=42$ and 48, but for $d=36$ (and also smaller values), the initial clump collapses to a black hole.
In Fig.~\ref{fig:collapse_lapse}, we can see the collapse of the lapse for
one of the quadrupole cases together with the corresponding dipolar case which does
not collapse.

\begin{figure*}
  \center
    \includegraphics[width=0.15\textwidth]{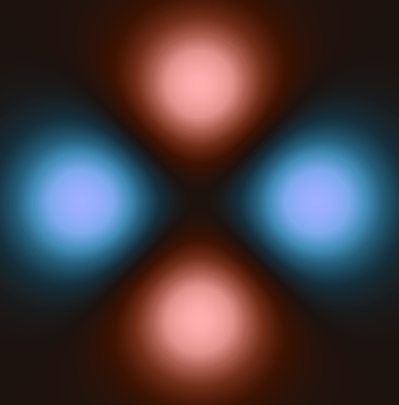}\,\dots
    \includegraphics[width=0.15\textwidth]{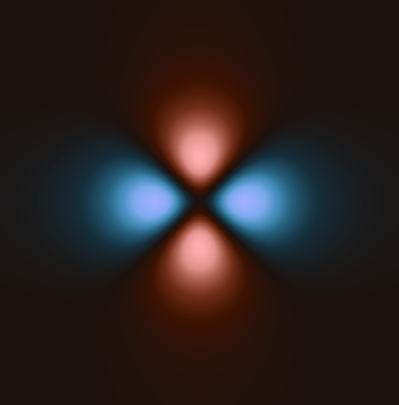}
    \includegraphics[width=0.15\textwidth]{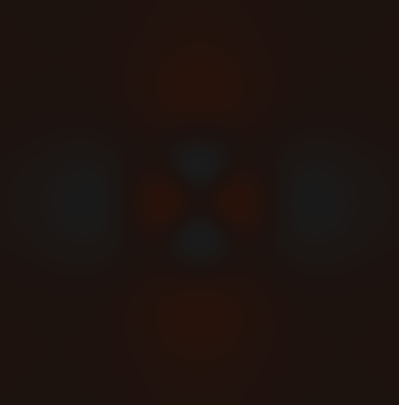}
    \includegraphics[width=0.15\textwidth]{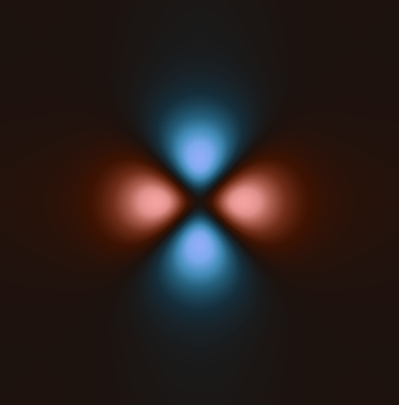}\\
		\includegraphics[width=0.8\textwidth]{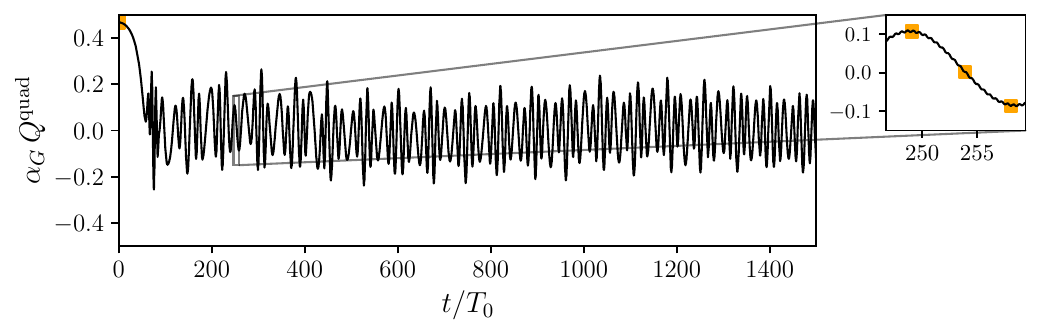}
  \caption{
    Charge density at $t=0$ of four boson stars with $\omega = 0.99$ and located
    at the vertices of a square with side length $d=42$ (top left panel) and three snapshots within
    half a charge-swapping period when the quadrupolar charge-swapping configuration
    has formed (top right panel). The 2D plots of the top panels correspond to the
    $xz$ plane. The color scheme uses
    blue for a positive density, red for a negative one and black for zero.
    The range of values spans from $-5.3\times10^{-5}$ to $5.3\times10^{-5}$.
    The box size is 50 for the top left panel, and 16 for the top right three boxes.
    We also show that the integral of the charge density over $z>|x|$, $Q^{\rm quad}$,
    as a function of $t$ (bottom panel). The points in the inset
    show the four snapshots of the top panels.
  }
  \label{fig:Q_quad}
\end{figure*}

Configurations that avoid collapse exhibit quadrupolar charge-swapping structures. Despite differing shapes, these structures show quantitative similarities to the dipolar case, following similar correlations between initial parameters $\omega$ and $d$, and the final solution properties. However, their lifetimes differ markedly from self-interacting quadrupoles without gravity ({\it i.e.,} quadrupolar Q-balls) \cite{Xie:2021glp}, with all cases persisting for at least $t = 5000T_0$. In Fig.~\ref{fig:Q_quad}, snapshots of charge density illustrate typical behavior, comparing initial star distributions at $t = 0$ with those within half a charge-swapping period later in the simulation. The bottom panel of Fig.~\ref{fig:Q_quad} shows $Q^{\rm quad}$, the total charge associated with one blob by integrating $n_\mu j^\mu$ over the $z > |x|$ region.

In this case, compared to the analogous plot for a dipole shown in Fig.~\ref{fig:Q_osc}, the initial relaxation process begins later. Specifically, because the four stars are initially far apart, the Noether charge $Q^{\mathrm{quad}}$ remains positive until approximately $t/T_0 \sim 50$, when the merger occurs. Following the merger, relaxation begins, and regular stable oscillations of the charge are observed after $t/T_0 \sim 100$. We have found that the scalar field ejected at the merger of dipolar and quadrupolar cases is essential to obtain a charge-swapping configuration.

It is natural to ask whether there are higher order multipolar boson stars. So far, all our preliminary attempts to use the method described in Section~\ref{sec:constraint} to find solutions have resulted in the configurations collapsing into black holes. As there is more mass involved in a small space, one can try to increase the distance $d$, but that leads to a decrease of the gravitational attraction. However, it is possible that octupolar boson stars may be constructed with a more systematical parameter scan or an improved preparation method.

\section{Self-interacting scalar}\label{sec:sifield}

Now, we include self-interactions for the U(1) scalar. In the flat space, they provide attractions for non-topological solitons and their molecular states to form. In the presence of gravity, scalar self-interactions can increase the compactness and the masses of boson stars. We will consider two types of interactions: a polynomial potential truncated to the sixth order and a soft logarithmic correction to the mass term, the charge-swapping Q-balls of which have been previously studied \cite{Copeland:2014qra, Xie:2021glp, Hou:2022jcd}.

\subsection{Sextic potential}
\label{sec:sexticpot}

The boson stars in the sextic potential model
\begin{equation}
    V(|\phi|)= |\phi|^2 - |\phi|^4 + \mathfrak{g} |\phi|^6\, 
\end{equation}
exist within the range
$\omega_{\rm min}<\omega<1$ with $\omega_{\rm min}$ some value that depends
on $\alpha_G$ and $\mathfrak{g}$ (see Fig.~\ref{fig:Mass} for
a sextic family of solutions with fixed $\mathfrak{g}$ 
and different $\alpha_G$). As already discussed, gravity regularizes the total energy of 
the solutions in the limit $\omega\to1$. In fact, $M\to0$ at such limit for all
cases (except $\alpha_G =0$). This is often known as the Newtonian limit.

According to our numerical experiments, starting from the Newtonian limit and
increasing the value of $f(r=0)$, there exist stable solutions
whenever 
\begin{equation}
\frac{dM}{d\omega}<0 
\end{equation}
until the turning point $\omega_{\rm min}$.
This leads to one or two stable branches according to the
value of $\alpha_G$; see Fig.~\ref{fig:Mass}. These observations are consistent with the conclusions drawn from a stability 
analysis of such solutions using catastrophe theory 
\cite{Tamaki:2010zz,Kleihaus:2011sx}.
After the turning point, all solutions are unstable regardless of the sign of $dM/d\omega$, some of them migrating
toward a stable boson star and others dissipating to infinity or collapsing to a 
black hole depending on the binding energy, which is similar to the free field and the quartic self-interaction cases
previously reported in the literature \cite{Seidel:1990jh,Balakrishna:1997ej}.

Compared to the free field case, the self-interacting solutions
are qualitatively very different when the gravitational
coupling is small, where they are close to their Q-ball counterparts.  
Nevertheless, these cases can give rise to very compact solutions (as $f(r=0)$
increases) that differ significantly from the Q-ball solutions and achieve
very high compactness, making them different from both the mini-boson stars and the corresponding Q-balls; see
Fig.~\ref{fig:numerical_application}. With this in mind, we 
have chosen the stars in Table \ref{tab:si_models} to prepare the charge-swapping 
configurations.
In particular, we restrict our study to superpositions of single 
star solutions that are stable on their own.
Of course, this does not guarantee that the complex boson star solution will be obtained, 
and neither ensures that the final solution will not be a black hole.
Since we have
fixed $\mathfrak{g}=1/2$ as our fiducial model, we only have two degrees of freedom
to construct the initial data: $\omega$ and $d$.
\begin{table}
  \centering
  \begin{tabular}{ll|ccccc}
    \hline\hline
    $\alpha_G\, (\times 8\pi)$ & $\omega$ & $M$ & $Q$ & $R_{99}$ & $\mathcal{C}$ & $E$  \\ \hline
    $0.001$  & $0.8$     & $863.5$     & $1009$  & $8.40$        & $0.00407$     & $843.9$\\
    $0.01$   & $0.8$     & $433.6$     & $491.8$ & $7.40$        & $0.0235$      & $375.4$\\
    $0.01$   & $0.75$    & $865.7$     & $1052$  & $8.02$        & $0.0417$      & $678.7$\\
    $0.01$   & $0.7$     & $1660$      & $2152$  & $9.03$        & $0.0728$      & $1113$\\
    $0.1$    & $0.7$     & $138.8$     & $159.1$ & $5.22$        & $0.1050$      & $67.42$\\
    \hline\hline
  \end{tabular}
  \caption{Global quantities of the boson stars used in the preparation of
    self-interacting charge-swapping configurations. All of them correspond
    to the model with $\mathfrak{g} = 1/2$. Notice that all of them correspond
    to low compactness stars. $E$ is calculated using the integral in
    Eq.~\eqref{eq:E} which tends to $M$ as $\alpha_G$ takes smaller values,
    consistent with the discussion below Eq.~\eqref{eq:komar2}.
    }
  \label{tab:si_models}
\end{table}

\begin{figure}
  \center
		\includegraphics[width=0.45\textwidth]{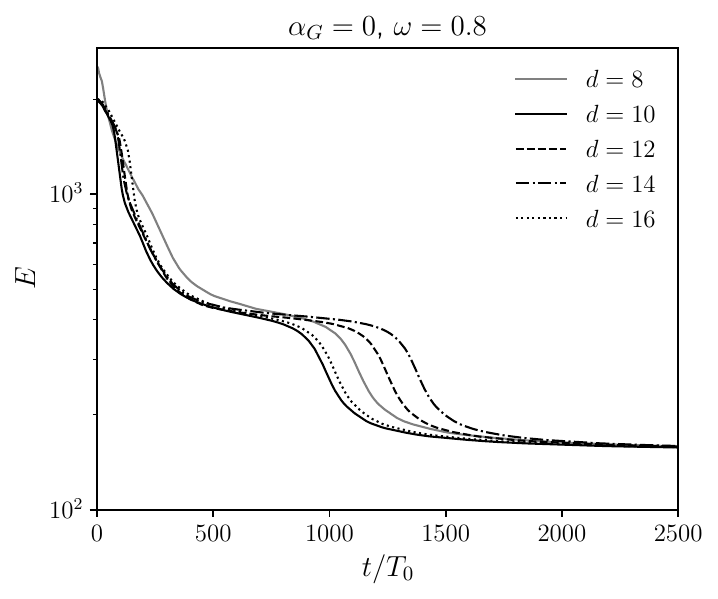}
	\caption{Flat spacetime simulations for the sextic model with $\mathfrak{g}=1/2$.
  We plot the energy as a function of time
  varying $d$ for fixed $\omega=0.8$. The $d=14$ case
  is consistent with that of Ref. \cite{Xie:2021glp}.}
	\label{fig:consistence}
\end{figure}

We shall explore the dipole superposition with initial
separations $d=\{8,10,12,14,16\}$. Before proceeding with the boson star analysis, as a sanity check,
we first re-produce the 3+1D results of 
charge-swapping Q-balls in Ref.~\cite{Xie:2021glp}, The resulting flat spacetime evolutions are shown in Fig.~\ref{fig:consistence}. 
We will see that the charge-swapping configurations, when gravity is included, share many properties with the flat spacetime limit analogue. In this sense the Fig.~\ref{fig:consistence} is representative of some of the properties that are obtained in general. There are four stages present when charge-swapping configurations are succesfully obtained:
An initial relaxation stage, which ends around $t=500T_0$ for the case presented in 
Fig.~\ref{fig:consistence}; a plateau stage of charge-swapping, ending around $t=1000T_0$;
a fast decay stage in which $Q^{\rm up}$ drops abruptly to zero and the energy
density also decreases but towards a finite spherically symmetric state;
a long-lived stage of the (highly perturbed) oscillaton \cite{Seidel:1991zh}, the oscillon in the absence of gravity (see Ref.~\cite{Zhang:2021xxa} for related real vector field solitons with cosmological applications).

Unlike the analysis conducted for the free field case, in all the scenarios explored with the sextic potential, we positioned the stars initially close together. For instance, in all the figures presented in this section, the maximum distance $d$ considered is 16. When comparing this to the radii of the stars listed in Table~\ref{tab:si_models}, it is evident that the stars are either ``touching'' or very close to each other initially in nearly all cases. Typically, the repulsive self-interaction from the quartic term is sufficient to prevent gravitational collapse, making the consideration of larger initial separations less relevant for the sextic potential compared to the free field scalar case. Consequently, the early stages of the relaxation process, including the merger, occur rapidly after the initial configuration of the stars.

We see from Fig.~\ref{fig:consistence} that the longest lifetime is given by $\omega = 0.8$ and $d=14$, consistent with that of Ref. \cite{Xie:2021glp}, among other results. These flat-space solutions have smaller radii than the 
free-field configurations with gravity discussed in Section~\ref{sec:freefield}. Nevertheless, for an easy comparison,  we
keep the same numerical setting as in the dipolar free field case with gravity.

\begin{figure}
	\center
		\includegraphics[width=0.45\textwidth]{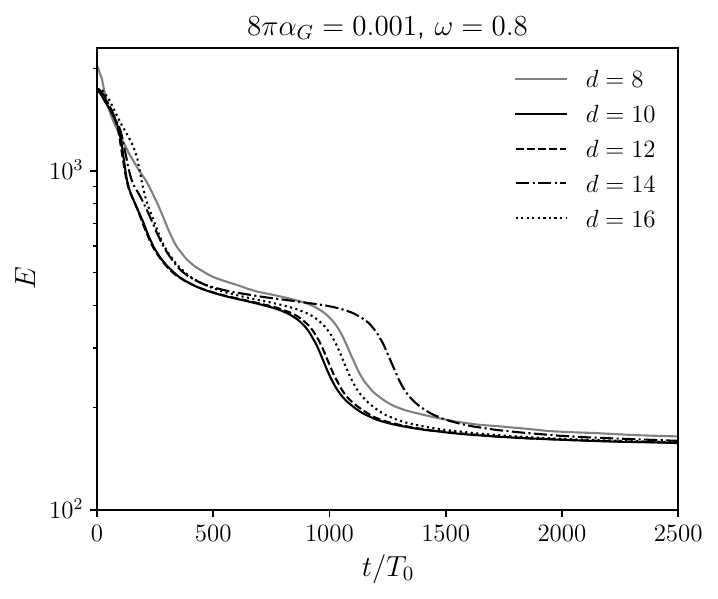}\\
    \includegraphics[width=0.45\textwidth]{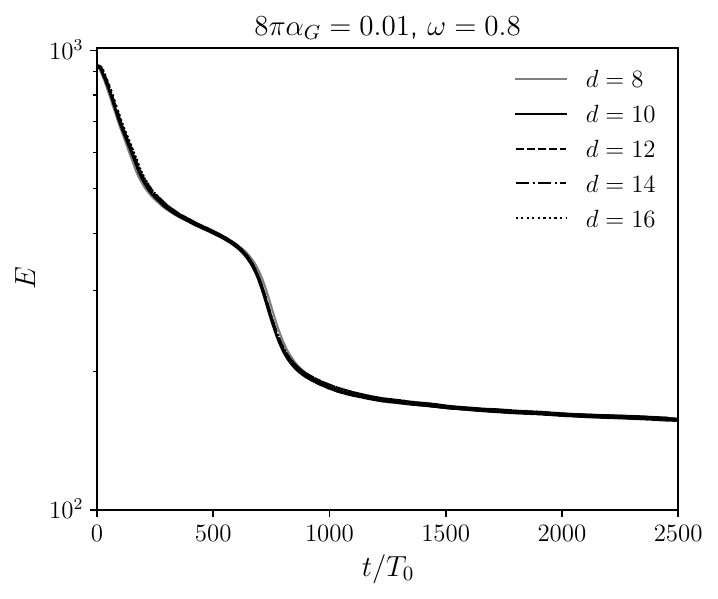}
	\caption{Almost the same setup as Fig.~\ref{fig:consistence} but with the gravitational coupling $\alpha_G$ switched on. 
  Again we vary $d$ and $\alpha_G$ for fixed $\omega=0.8$.}
	\label{fig:results_E1}
\end{figure}

Next, we switch on $\alpha_G$
to investigate the effect of introducing the gravitational interaction.
To monitor the evolution in this case, although not perfect as it is now not really conserved, we will still use the $E$ integral as a proxy. In all the cases presented in this section, we observe that the
envelope of the Noether charge in the upper hemisphere always 
follows a similar qualitative behavior to that of the $E(t)$ curve. 
In Fig.~\ref{fig:results_E1}, we show the cases with frequency $\omega=0.8$
and gravitational coupling $8\pi\alpha_G = 0.001$ and $0.01$.
This corresponds to the two first rows in Table~\ref{tab:si_models}.
For $8\pi\alpha_G=0.001$, $\omega=0.8$ is the point where the spherical boson stars start to deviate from the
Q-ball counterparts (see Fig.~\ref{fig:Mass}). The same is also noticeable in the top panel of Fig.~\ref{fig:results_E1} for the charge-swapping configurations, where we see that the lifetimes are slightly shortened in the presence of gravity for all the $d$'s listed. For a smaller $\alpha_G$, the system basically reduces to the Q-ball case. For $8\pi\alpha_G=0.01$ in the bottom panel, interestingly, we see that the differences between the $E$ curves are negligible 
for different $d$'s, due to the significantly reduced lifetimes. This means that the gravitational interaction is already dominating for $8\pi\alpha_G=0.01$, and the stable end state corresponds to an oscillaton as expected. To confirm that the final stage is indeed an oscillaton, we examine the quantity  $Q^{\mathrm{up}}$. In regions where this quantity approaches zero, we analyze the phase portraits of the complex scalar field, specifically plotting $\Re(\phi)$ versus $\Im(\phi)$. During the oscillaton stage, the phase portraits exhibit oscillations along a single linear direction.

In these cases presented in Fig.~\ref{fig:results_E1}, as well as in the $\alpha_G\to0$ limit another relevant mechanism takes place in the formation of charge-swapping states. When self-interaction is significant this mechanism is characterized by a net attraction force arising from the nonlinear scalar field, which is specific to charge-swapping systems and has been detailed in Ref.~\cite{Copeland:2014qra} for Q-balls. In our configurations, when the quartic term of the scalar potential is large and the effective gravitational constant is therefore small, this property predominates in the attraction between the two blobs, facilitating the formation of bound states. In the free-field scenario, this effect is absent, with the gravitational interaction being the sole force that holds the stars together. As we will demonstrate, when we increase the value of $\alpha_G$ for the sextic potential case under examination in this section, the interaction that initially facilitated the formation of bound states becomes subdominant, allowing gravity to dominate the dynamics of the system.

Now, we turn to the cases where the initial boson stars are not similar to the
Q-ball analogue with the same value of $\omega$. This corresponds to the three last rows in
Table~\ref{tab:si_models}. The two configurations with 
$\omega = 0.7$ are beyond the so-called thin-wall limit of the corresponding 
Q-ball model, and in particular the star with $8\pi\alpha_G = 0.1$ has a
larger compactness than any of the stable mini-boson stars, 
$\max \mathcal{C}_{\rm mini} = 0.08$ \cite{Amaro-Seoane:2010pks}.
We find that the charge-swapping 
configurations can be created for the cases with $8\pi\alpha_G = 0.01$
and $\omega = 0.7$ or 0.75, but they collapse in the $8\pi\alpha_G= 0.1$
case. 

\begin{figure*}
	\begin{centering}
		\includegraphics[width=0.435\textwidth]{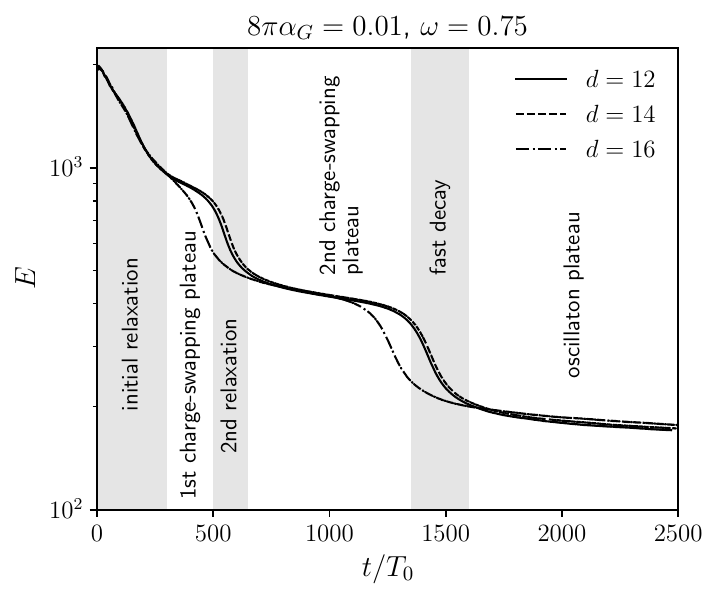}\quad\includegraphics[width=0.45\textwidth]{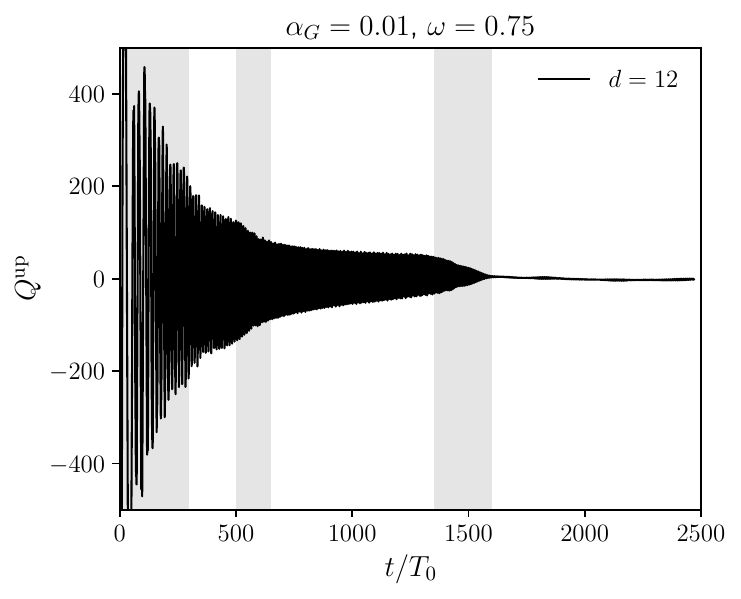}
	\end{centering}
	\caption{Varying $d$ for fixed $\omega=0.75$. In the right panel we show the Noether charge
  in the $z>0$ region, $Q^{\rm up}$, only for the $d=12$ case. For reference we mark the different stages of the evolution for this case in both plots.}
	\label{fig:results_E3}
\end{figure*}

\begin{figure}
	\begin{centering}
		\includegraphics[width=0.45\textwidth]{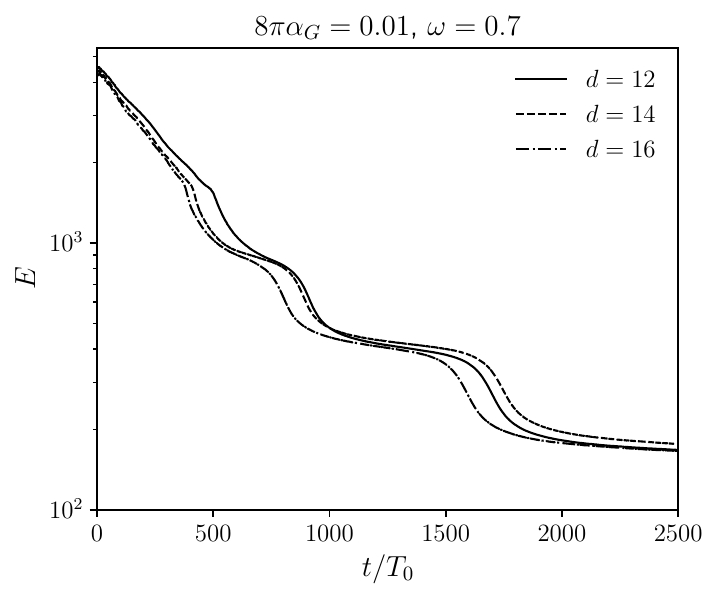}
	\end{centering}
	\caption{Varying $d$ for fixed $\omega=0.7$ and $8\pi\alpha_G = 0.01$. This self-gravitating configuration was prepared using boson stars with initial frequencies smaller than those allowed by the flat spacetime Q-ball thin-wall limit.
  }
	\label{fig:results_E2}
\end{figure}

The evolution of $E$ for $\omega = 0.75$ and $8\pi \alpha_G=0.01$ for three
different $d$'s
together with the total Noether charge in the $z>0$ region for $d=16$
is shown in Fig.~\ref{fig:results_E3}.
We can see how  the charge is radiated away in the final stage. 
More interesting is the fact that for these cases, as well as for the cases in Fig.~\ref{fig:results_E2}, the 
configuration transits through two charge-swapping plateaus 
with different $E$ and $Q^{\rm up}$ before 
migrating to the oscillaton configuration. This resembles the cascading of the energy
for excited oscillons \cite{Wang:2022rhk}. However, at the level of the quantities we have observed, a direct connection between the two phenomena has yet to be established.

The case beyond the thin-wall limit with $8\pi\alpha_G = 0.01$ is shown in Fig.~\ref{fig:results_E2}. For $8\pi\alpha_G = 0.1$, similar setups would lead to gravitational collapse, without stabilizing first to a finite value of $E$ or 
$Q^{\rm up}$.

The last comment we shall make for the case of the sextic potential is 
that isolated boson stars differ from Q-balls not only in cases near 
or beyond the thin-wall limit but also in the region where gravity 
regularizes the solutions near $\omega = 1$. Since  in this limit the total
mass of the solution tends to zero, the curve of $M$ vs $\omega$
(or radius) is ``forced'' to form a stable Newtonian solution branch.
In this branch, the stationary solutions of the sextic potential 
converge to those of the quadratic potential in the limit, due to the fact that $\phi\to0$ in this limit; see \cite{Siemonsen:2020hcg} 
for a systematical study of several
kinds of self-interactions. 
It is therefore expected that configurations prepared using this 
branch of solutions will lead to the same results as those 
presented in Section~\ref{sec:freefield}.

\subsection{Logarithmic potential}
\label{sec:logpot}

Polynomial potentials naturally arise as effective potentials when integrating out weakly coupled heavy degrees of freedom. On the other hand, when including quantum corrections, the potential often picks up some logarithmic dependence. These potentials are widely studied for the case of Q-balls \cite{Enqvist:1997si,Enqvist:1998en,Enqvist:1999mv,Multamaki:1999an,Enqvist:2000gq,Kasuya:2000wx}, and  
capable of creating very long-lived charge-swapping 
Q-balls \cite{Copeland:2014qra, Hou:2022jcd}. In this subsection, we consider a complex scalar field minimally coupled with Einstein gravity with a potential of the kind,
\begin{equation}
  V(|\Phi|) = \mu^2|\Phi|^2\left(1+K\ln\frac{\, |\Phi|^2}{\mathcal{M}^2}\right) \, ,
\end{equation}
where $K$ is a negative coefficient that we will fix to $K = - 0.1$ in the following
and the parameters $M$ and $\mu$ are scales
that can be absorbed by redefinitions.

As mentioned previously, it is also possible to obtain a dimensionless
action in terms of dimensionless variables and parameters.
To this end, we simply choose $x^\mu = \mu \tilde{x}^\mu$ and $\phi = \Phi/\mathcal{M}$.
In doing so, we again gain control over the gravitational interaction,
defining in this case $\alpha_G=\mathcal{M}^2G$ while rescaling the Ricci scalar as in the sextic potential case. Thus, we can study the action
$\tilde{S} = \mathcal{M}^2 S/\mu^2$ instead, with the dimensionless potential given by 
\begin{equation}
V(|\phi|) = |\phi|^2\left(1+K\ln|\phi|^2\right) \,.
\end{equation}
Once again the equations of motion for the gravitational and scalar fields are given by
Eqs.~\eqref{eq:einstein} and \eqref{eq:kg}.

An interesting property of this model is that in the Minkowski limit ($\alpha_G \to 0$), the equation of motion for $\phi$
\begin{equation}
  -\frac{\partial^2 \phi}{\partial t^2} + \Delta_3 \phi = \left[1+K(1+\ln|\phi|^2)\right] \phi \, ,
\end{equation}
admits an exact spherical solution
\begin{equation}\label{eq:exactphi}
  \phi = \exp\left(\frac{\omega^2-1}{2K}+1\right) \exp\left(\frac{Kr^2}{2}\right) \exp(-i\omega t)\, .
\end{equation}
Then, in order to build complex boson star solutions, we take this as initial guess and slowly increase
the value of $\alpha_G$, solving once again for the function $f$ of the ansatz
$\phi = f(r) e^{-i\omega t}$ and the metric coefficients $\alpha$ and $\Psi$ using
Eqs.~\eqref{eq:BS_E} with the logarithmic potential. 
Then we proceed to prepare initial data for
charge-swapping configurations following the same procedure as described in Section~\ref{sec:constraint}.
The opposite limit $\alpha_G\to\infty$ needs to be considered with more
care, but is beyond the scope of this work.

\begin{figure*}
	\begin{centering}
		\includegraphics[width=0.45\textwidth]{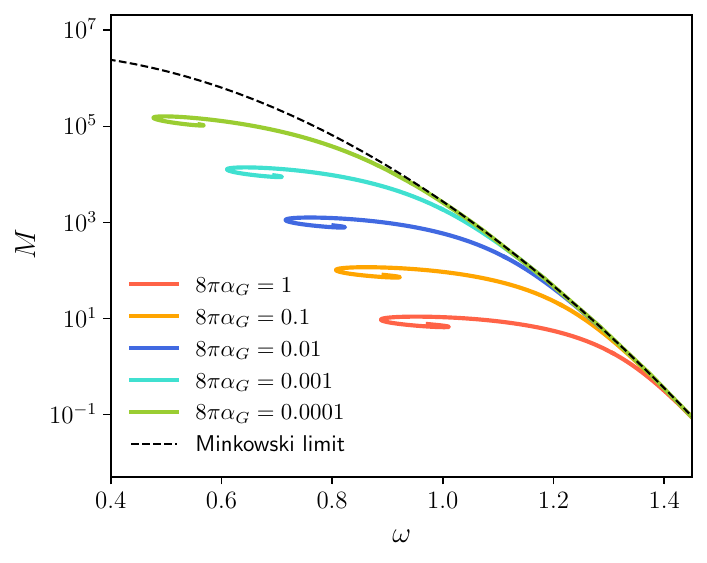} \quad\includegraphics[width=0.45\textwidth]{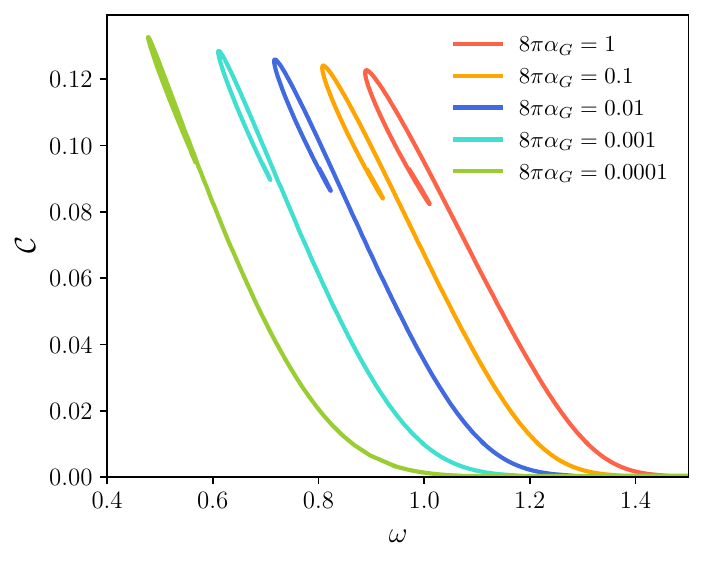}
	\end{centering}
	\caption{Sequence of spherical boson stars for the logarithmic potential with $K = - 0.1$, $M$ being the dimensionless mass 
          and $\mathcal{C}$ being the compactness. We see that $M$ converges to the Q-ball
          solution in Eq.~\eqref{eq:exactmass} as $\alpha_G\to 0$. In particular, 
          for small values of $\phi$ (low mass)
          the solutions approach the Minkowski case, where $M$ decreases exponentially with frequency $\omega$.
  }
	\label{fig:logMass}
\end{figure*}

Sequences of the spherical boson stars for the logarithmic potential are presented in 
Fig.~\ref{fig:logMass} for five different values of the gravitational
coupling constant.
In this figure, we have included also the $\alpha_G\to0$ limit, which can be obtained by
substituting Eq.~\eqref{eq:exactphi} into Eqs.~\eqref{eq:E} and \eqref{eq:Q}, giving rise to the following expressions for $E=M$ and $Q$,
\begin{align}\label{eq:exactmass}
M&=\left(-\frac{\pi}{K}\right)^{3/2}(2\omega^2-K)\exp\left[\frac{\omega^2-1}{K} +2\right] \, ,
\\
\label{eq:exactcharge}
Q&=2\left(-\frac{\pi}{K}\right)^{3/2}\omega\exp\left[\frac{\omega^2-1}{K} +2\right] \, .
\end{align}

For all the boson star families plotted in Fig.~\ref{fig:logMass}, we find that, interestingly, the frequency $\omega$ of the critical mass increases with $\alpha_G$.
To the best of our knowledge, no such boson stars have been previously constructed in the literature (see \cite{Choi:2019mva} where a different logarithmic potential was explored).
Consequently, no results are available on the stability of such spherical boson stars.
We can anticipate that beyond the critical mass point the configuration is unstable,
as is the case for most (if not all) families of spherical boson stars available in the literature.
On the other hand, configurations to the right of the respective maxima in Fig.~\ref{fig:logMass}
are potentially stable, so we use spherical boson stars with $\omega = 1.2$ to prepare charge-swapping configurations. In Table \ref{tab:log_models},
we display the classical observables of the logarithmic boson star solutions we will use.
\begin{table}
  \centering
  \begin{tabular}{l|ccccc}
    \hline\hline
    $\alpha_G\, (\times 8\pi)$ & $M$         & $Q$     & $R_{99}$       & $\mathcal{C}$     & $E$    \\ \hline
    $0.0001$                   & $47.54$     & $38.29$ & $7.76$        & $2.4\times10^{-5}$ & $46.78$\\
    $0.001$                    & $46.97$     & $37.82$ & $7.71$        & $2.4\times10^{-4}$ & $46.15$\\
    $0.01$                     & $42.15$     & $33.88$ & $7.68$        & $2.2\times10^{-3}$ & $40.85$\\
    $0.1$                      & $22.98$     & $18.29$ & $7.22$        & $0.0127$          & $20.66$\\
    $1$                        & $5.512$     & $4.287$ & $6.38$        & $0.0343$           & $4.176$\\
    \hline\hline
  \end{tabular}
  \caption{Mass, charge, radius and other global properties of
    the boson stars used in the preparation of
    logarithmic charge-swapping configurations. All of them correspond
    to the model with $K=-0.1$ and have a frequency $\omega = 1.2$.
    This corresponds to the gravitational
    generalization of the fiducial model explored in \cite{Hou:2022jcd}.
    The value of $E$ is calculated using the integral in
    Eq.~\eqref{eq:E}, which tends to the value of $M$ when $\alpha_G$ is small.
    }
  \label{tab:log_models}
\end{table}

Interestingly, all the presumably stable configurations shown in Table \ref{tab:log_models}
have a positive binding energy 
\begin{equation}
    E_{\rm bind} = M - Q \, .
\end{equation} 
This is different to all the other
cases explored in this work, Tables \ref{tab:ff_models} and \ref{tab:si_models}.
This can be seen from the exact expression for $M$ and $Q$ (see Eqs.~\eqref{eq:exactmass} and \eqref{eq:exactcharge}). 
Yet, these configurations survive even against finite perturbations for a very long time.
We have explored the 5 configurations in the Table \ref{tab:log_models} together with the
fiducial model in \cite{Hou:2022jcd} and have not seen any of these configurations decay within our simulation limits. Thus, they are perfectly suitable for preparing
charge-swapping configurations.

We take a dipole superposition with $d = 2$ for the five cases in Table \ref{tab:log_models},
and evolve them using a modified scalar field thorn in a mesh with the same specifications
as those of the sextic potential case. The configurations are of the same size. We find that all
of them form long-lived charge-swapping configurations except for the case with $8\pi\alpha_G = 1$,
which promptly collapses to a black hole before any charge-swapping takes place.
All the other cases are very long-lived and survive at least until $t = 60000$.

\begin{figure}
	\center
		\includegraphics[width=0.45\textwidth]{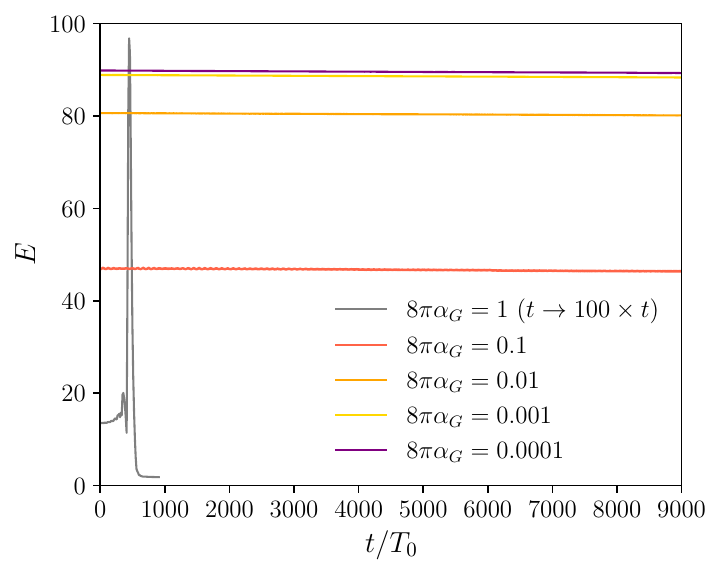}
	        \caption{Charge-swapping configurations for the logarithmic potential with $\omega = 1.2$, prepared
                  from a distance $d=2$ for different values of the gravitational coupling constant.
                  The $t$ axis for the $8\pi\alpha_G = 1$ case is scaled by a factor of 100 in order
                  to visualize its collapse.}
	\label{fig:E_log}
\end{figure}
\begin{figure}
	\center
		\includegraphics[width=0.45\textwidth]{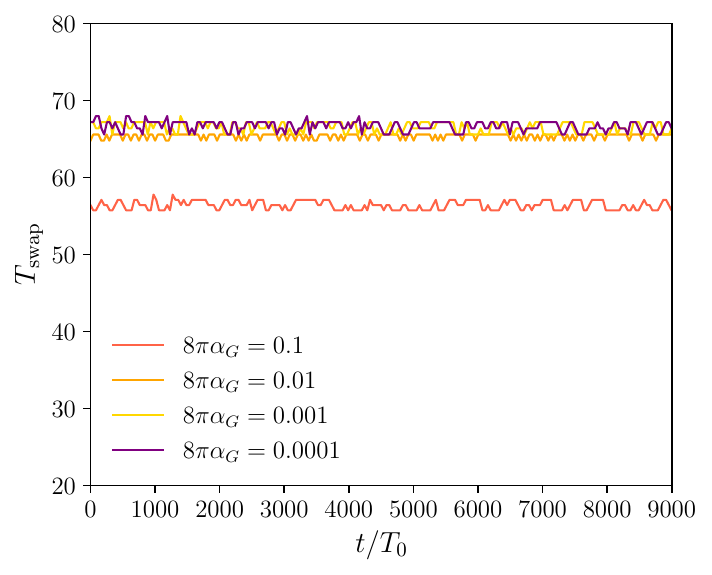}
	\caption{Charge-swapping period $T_{\rm swap}$ for the logarithmic charge-swapping boson stars in Fig.~\ref{fig:E_log}.}
	\label{fig:T_log}
\end{figure}

As in the sextic potential case, here we also prepared configurations that are close to each other at $t=0$, so in this sense we are expected to start with states close to the charge-swapping configuration. Interestingly, similar to the charge-swapping Q-ball case with the logarithmic potential,
there is no relaxation process that radiates away excess of the scalar field, as can be
confirmed in Fig.~\ref{fig:E_log}. Also, the
charge-swapping process begins very early in the simulations.
The latter is reflected in the fact that the charge-swapping period $T_{\rm swap}$ is clearly
defined from the beginning of the evolution, as can be seen in Fig.~\ref{fig:T_log}: $T_{\rm swap}$ is defined as twice the time elapsed between two contiguous zero points of the Noether charge in the upper half space $Q^{\rm up}$.

\section{Anisotropic distribution of satellite galaxies}\label{sec:sat_galaxies}

\begin{figure}
  \centering
	  \includegraphics[width=0.42\textwidth]{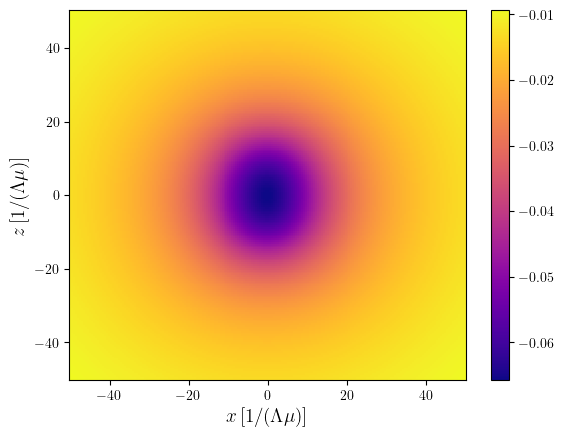}\\          \includegraphics[width=0.35\textwidth]{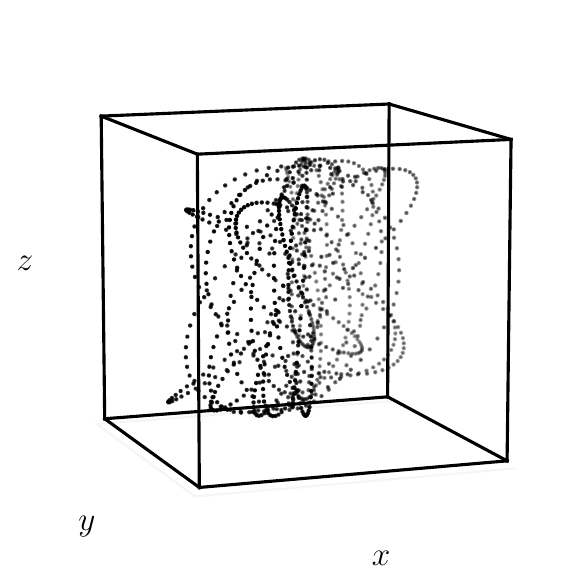}
	\caption{Newtonian gravitational potential $\Lambda U$ of the dipolar charge-swapping configuration with $\omega =0.98$ (top) and motion of a single random ``particle''
          up to $t=12\tau_s$ using 1000 steps of time (bottom). $\tau_s$ is the time for a ``particle'' in the equatorial plane to
complete a circular orbit.
  }
	\label{fig:potential}
\end{figure}

In the last two sections, we have explored generic features of complex boson stars, particularly their lifetimes and the charge distributions. As we have seen, these ``molecular states'' of boson stars are often quite stable and with distinct charge-swapping patterns. The sizes of these objects are not explicitly specified in these simulations, and in fact, $\mu^{-1}$ is essentially used as a base length, setting their characteristic sizes. In this section, we shall discuss one specific application of these objects in galactic scales. 

For the lowest order/dipolar charge-swapping configuration, its energy density distribution is almost constant with time, mostly spherical but distorted with an appreciable dipolar contribution, and the Noether density has a dipolar structure that is oscillating. This resembles the energy density
distribution of a scalar field configuration used previously as an alternative explanation \cite{Solis-Lopez:2019lvz}
to solve the problem of the
anisotropic distribution of satellite galaxies 
observed in the Milky Way, M31
and Centaurus A.
Other explanations for the plane of satellite galaxies problem include using baryonic effects, combined
gravitational distribution effects \cite{Sawala:2022xom} and 
formation of satellite galaxies within the scalar field dark matter model \cite{Park:2022lel}; see \cite{Pawlowski:2021ipt} for a review. 

The plane of satellite galaxies problem arises from the distribution and motion of these galaxies in planes that are perpendicular to the plane of their host galaxy. This is unnatural to explain in the cold dark matter
paradigm, where the simulations predict that satellites should be isotropically
distributed \cite{Shaya:2013xna,Pawlowski:2018sys}.
The authors in \cite{Solis-Lopez:2019lvz} address this problem
using multi-state solutions of a system of
self-gravitating scalar fields in the Newtonian limit, composed
of a spherical state and a dipolar state, which allow them to model dark matter halos and study the
motion of test particles on top of the generated gravitational potential. They found
that a anisotropic tri-axial Navarro-Frenk-White halo does not lead the orbital
angular momenta of the satellites to align with the equator, but the anisotropic
(multi-state) scalar field halo does accommodate the particles in orbital planes
close to the galactic poles. This is made possible because of the specific morphology
of the dipolar contribution, which is produced by a scalar field
anti-symmetric with respect to the galactic plane.

\begin{figure*}
	\begin{centering}
	  \includegraphics[width=0.4\textwidth]{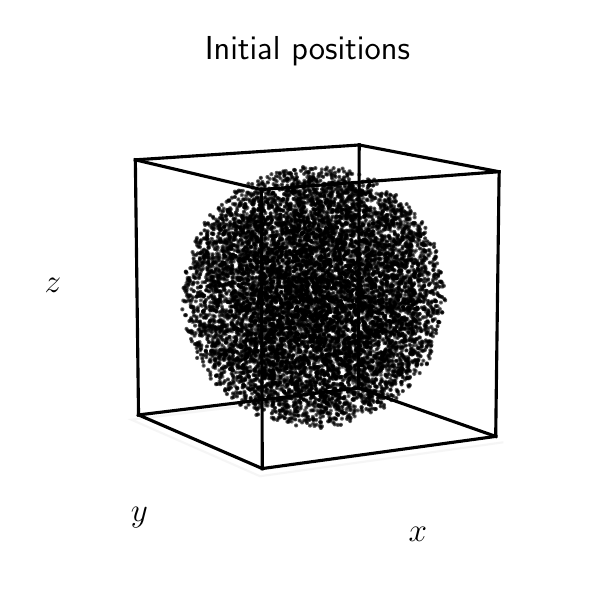}
          \includegraphics[width=0.4\textwidth]{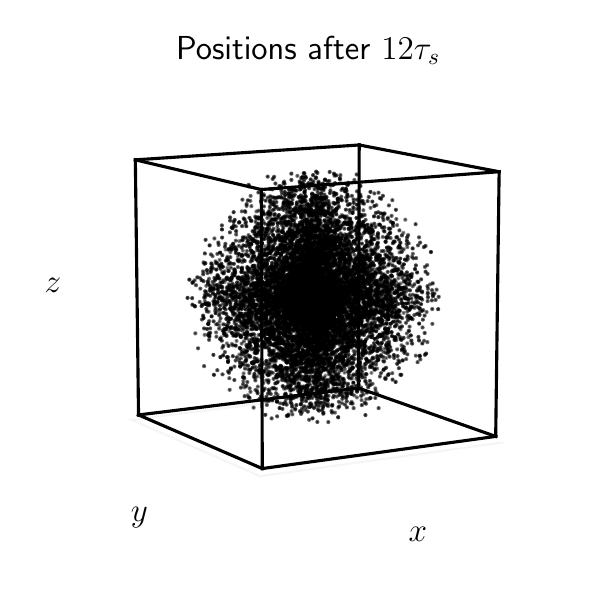}
	\end{centering}
	\caption{``Particles'' in the configuration space moving in
          the (Newtonian) gravitational potential of a
          dipolar charge-swapping configuration in a free scalar model. The length of the cube's side is $2\times300$ kpc.
          The configuration that gives rise to the dark matter halo, 
          on top of which ``particles'' are moving is a dilute dipolar configuration with
           $\omega = 0.98$. For
          reference, the radius of the constituent boson star in this scenario
          is $\sim600$ kpc and the formed complex boson star is about
          half the size of it.
  }
	\label{fig:particle_positions}
\end{figure*}

The self-gravitating charge-swapping configurations in the case of a
free scalar field or the scalar field with a logarithmic potential are
extremely long-lived and have a morphology similar to the monopole
dominating configuration of \cite{Solis-Lopez:2019lvz} (close to spherical symmetry
but with a sizable dipolar contribution). More importantly, their
constituent boson stars are close to the Newtonian regime, so these dipolar configurations are very dilute, fully compatible with cosmological
constraints.
In the following, we explore the possibility of utilizing charge-swapping configurations, which, as nonspherical halo distributions, could lead to the desired anisotropic distribution of satellite galaxies.
For definiteness, we will focus on charge-swapping dipolar configurations whose constituents are mini-boson stars with $\omega = 0.98$.

In the non-relativistic limit, the Einstein-Klein-Gordon system Eqs.~\eqref{eq:einstein}
and \eqref{eq:kg} reduce to the Sch\"odinger-Poisson system which is more tractable and also has additional re-scaling properties \cite{jaramillo19}. This approximation is valid
when the scalar field
is non-relativistic which in particular implies $\phi\ll1$. So even in the presence of higher order polynomial self-interactions, the scalar equation of motion reduces to
\begin{align}
  i\partial_t\Psi_{\rm w} &= -\frac{1}{2}\nabla^2\Psi_{\rm w} + U\Psi_{\rm w} \, , \\
  \nabla^2 U              &= 4\pi\alpha_G|\Psi_{\rm w}|^2 \, .
\end{align}
where
\begin{equation}\label{eq:wavefunction}
  \Psi_{\rm w} = \sqrt{2} \exp(i t)\phi \, ,
\end{equation}
$U$ is the Newtonian gravitational potential, which can be extracted by comparing Eq.~\eqref{eq:3+1} with
\begin{equation}\label{eq:nr_metric}
  ds^2 = -(1+2U)dt^2 + (1-2U) dr^2 + r^2d\Omega^2 \, ,
\end{equation}
A consistency condition is that $U\ll1$. We assume that the weak field and low energy regime
is valid in the late universe. Boson stars are known to be in this limit when $\omega$ is close
to 1, where the compactness approaches 0.

Unlike the Einstein-Klein-Gordon system, the Schr\"odinger-Poisson equations have\footnote{Subject to the restrictions $\phi\ll1$ and $U\ll1$. See Ref. \cite{Annulli:2020lyc} for the post-Newtonian expansion of the
Einstein-Klein-Gordon system.} the following scaling invariance:
\begin{equation}
(\Psi_{\rm w},U,x^i,t)   \rightarrow   (\Lambda^2\Psi_{\rm w},\Lambda^2U,\Lambda^{-1}x^i,\Lambda^{-2}t) \, ,
\end{equation}
for any value of the real scaling constant $\Lambda$.
This allows to obtain the full sequence of Newtonian boson stars
once a single solution has been found, unlike the relativistic analogue, where the family of solutions, say, the one presented in Fig.~\ref{fig:Mass}, must be constructed numerically solving the equations at each point. 
This rescaling is independent of the rescaling used to obtain dimensionless quantities, so to recover the physical system, as in the relativistic case, we must choose a value of $\mu$ as explained in Section \ref{sec:units}. We have previously obtained the metrics in the full general relativistic formulation. For this
application, it is sufficient to extract the non-relativistic limit 
from Eq.~\eqref{eq:wavefunction} and Eq.~\eqref{eq:nr_metric}. 

We choose to perform the analysis at a certain representative space slice with $t=4000T_0$
of the dipolar configuration with $\omega = 0.98$ well within the charge-swapping
stage. However, the results are insensitive to the specific time period chosen within this stage. To extract the 2D data at such time we use the \texttt{Kuibit} tool \cite{Bozzola:2021hus}.
Based on the consistency statistical tests performed in \cite{Solis-Lopez:2019lvz},
we choose the values of $\mu$ and $\Lambda$, which completely fix the scale of the system, to be such that the size of the configuration, which has an equatorial radius of about 16 units, corresponds to a physical size of 300 kpc (radius encompassing the 11 classical satellites of the Milky Way \cite{Pawlowski:2019bar}), together with the condition that the circular velocity of the stars in the galactic disk at a point 30kpc away from the galactic center is 100 km/s, implying similar contributions from the enclosed mass at radius r=30kpc of dark matter and baryonic matter.
These two conditions lead to the values $\mu\hbar \sim 10^{-25}\text{eV}$ and $\Lambda = 0.01$. The precise determination of these two parameters given the rotation curve of the Milky Way and a model for the galactic bulge and disc deserves a statistical analysis which is beyond the scope of this paper.

Next, we study the motion of $10^4$ ``particles'' moving the gravitational field of $U$. After some time, the distribution of the ``particles'' reflects the
probability of finding (idealized) satellite galaxies in the halo. 
To that end, we randomly place ``particles'' inside a sphere of radius
$R=16/(\Lambda\mu)$, corresponding to $R=300 \mathrm{kpc}$ (see the left panel of
Fig.~\ref{fig:particle_positions}), and with velocities in
random directions and magnitudes smaller than 1/4
of the escape velocity of a ``particle'' located in the equatorial plane at a
distance $R$ from the center of the galaxy: $v<v_{\rm max} = \sqrt{U(R,0,0)/8}$.

The ``particles'' are left to evolve. In Fig.~\ref{fig:potential}, we show the gravitational potential
and the motion of a single ``particle'' in it,
in units of $\mu$ and $\Lambda$, which specify the galactic sizes.
This is possible because the equations of motion of the ``particles'' inherit the two scaling invariances of the Shr\"odinger-Poisson equations mentioned above. Similar to \cite{Solis-Lopez:2019lvz}, we choose to evolve up to $t=12\tau_s$, where $\tau_s$ is defined as the time it takes for a particle in the equatorial plane to
complete a circular orbit. After this, we restore units of every involved physical quantity.

\begin{figure*}
	\begin{centering}
	  \includegraphics[width=0.2\textwidth]{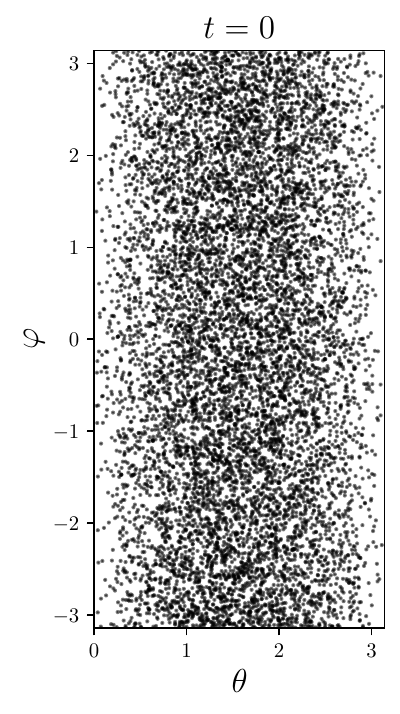}
          \includegraphics[width=0.2\textwidth]{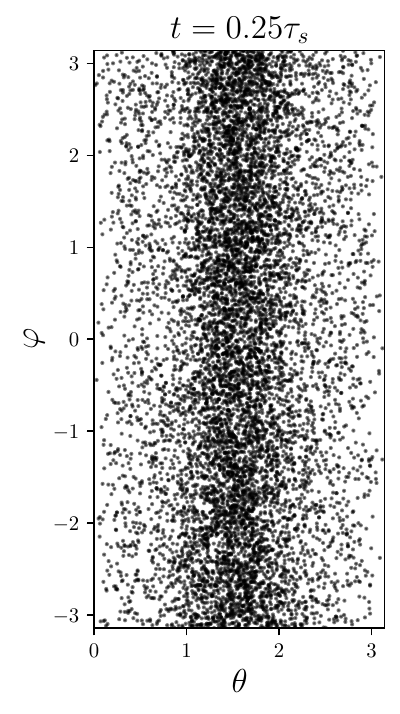}
          \includegraphics[width=0.2\textwidth]{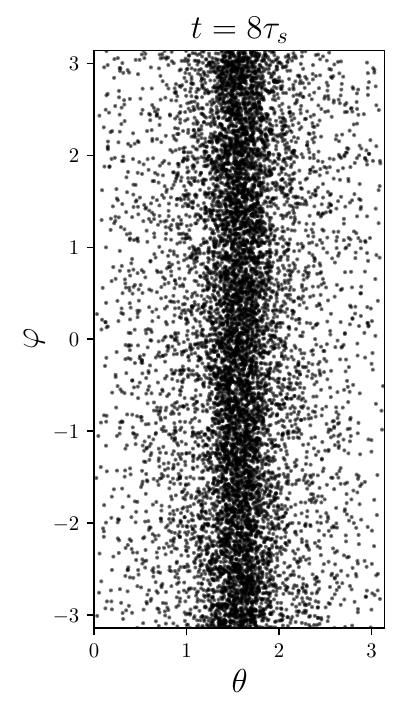}
          \includegraphics[width=0.2\textwidth]{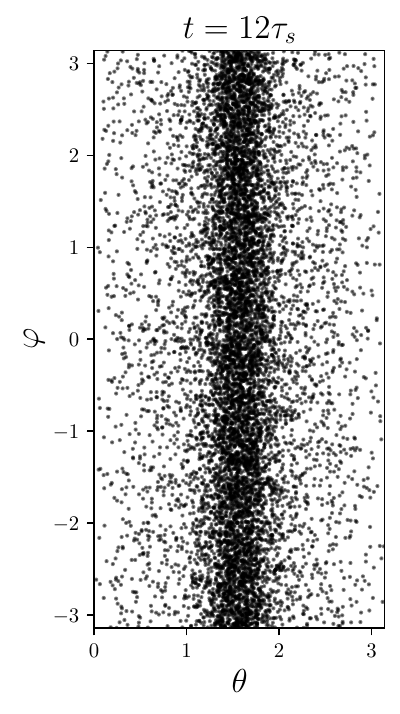}
	\end{centering}
	\caption{Angular distribution of the angular momenta of $10^4$ ``particles'' for the
          fiducial dipolar charge-swapping configuration. The initial distribution is isotropic,
          and it quickly relaxes to a configuration where the tangent of the orbital poles are
          oriented toward the galactic plane $\theta=\pi/2$ . We only include particles
          within $\Lambda\mu R\in (0.16,16)$ which takes into account ``particles''
          with $R>30$ kpc, the size of the Milky Way stellar disc.
  }
	\label{fig:orbital_poles}
\end{figure*}

\begin{figure*}
	\centering
          \includegraphics[width=0.8\textwidth]{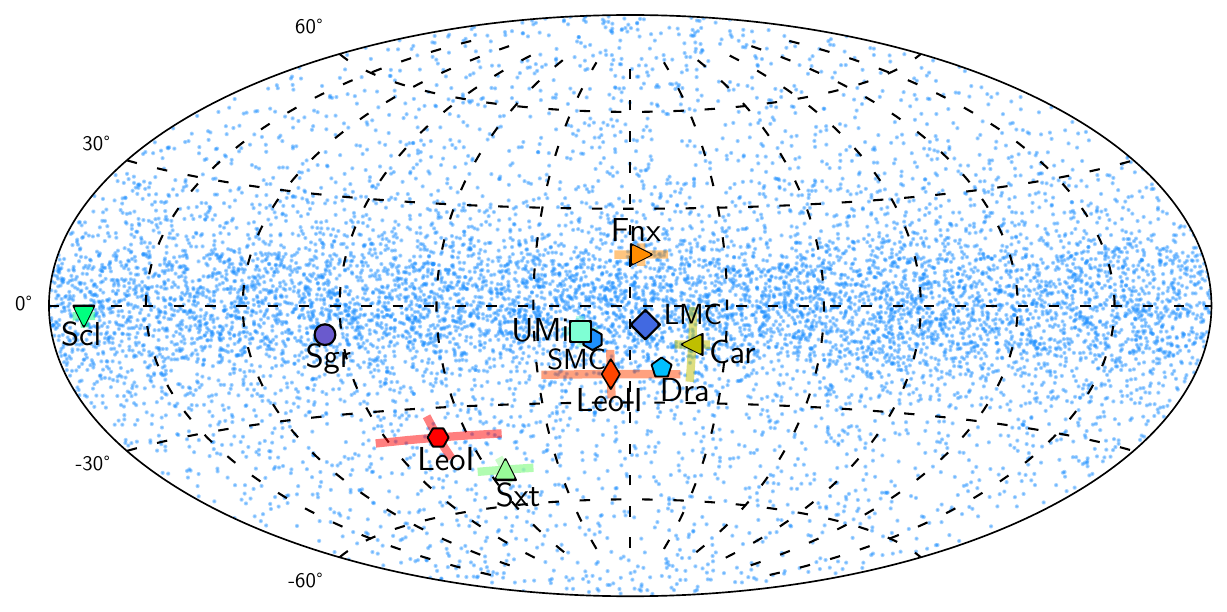}
	\caption{Orbital poles and their uncertainty of the 11 Milky Way
          classical satellites.
          Here the angles correspond to the Galactic longitude and latitude.
          The thick color lines correspond to the error bars in both angles for each
          satellite, arising from the uncertainty in determining the
          velocity components of the galaxies.
          The blue dots in the background correspond to the $10^4$ ``particles'' at
          $t=12\tau$ of Fig.~\ref{fig:orbital_poles}.
  }
	\label{fig:classical_satellites}
\end{figure*}

From Fig.~\ref{fig:particle_positions} and similar plots and projections
at different times,
we see that after several $\tau_s$, the distribution tends to a stationary
non-spherical distribution in space, and the distribution of the orientation of the
satellite orbits is also stationary and 
anisotropic. This can be seen in Fig.~\ref{fig:orbital_poles}.
More importantly, we notice that
the orbital angular momenta (known as the orbital pole in
the astronomy literature) of the satellites, {\it i.e.,} the ``particles'' located
between $30$ and $300$ kpc, are concentrated around the equator. Other choices
of the maximum allowed value for the random magnitude of the velocity lead to
similar results as long as it does not exceed 1/2 the escape velocity of the particle located in the equatorial plane at a distance $R$ from the center. For a comparison, we have plotted the orbital poles of the
11 Milky Way classical satellites obtained from Ref.~\cite{Pawlowski:2019bar}
together with the particle orbital poles
at $t=12\tau_s$ in Fig.~\ref{fig:classical_satellites}.

Let us comment on the robustness of the charge-swapping configurations for
modeling dark matter halos.
In our general relativistic numerical simulations, we find that the free-field
charge-swapping configurations can be formed from quite generic initial setups and are attractor solutions in the dynamical evolution. 
Also, as mentioned in the introduction, they may initially be formed from fragmentation of some homogeneous condensate in the early universe. 
They live for extremely long times and are sufficiently stable in terms of modeling a galactic halo. To see this, note that for our case we have that the dimensionless timescale is $\Lambda^2\mu\tau_s\sim600$, calculated from $U$ according to its definition, which is much smaller than the (minimum) lifetime of the free-field charge-swapping configurations explored in Section \ref{sec:freefield}. For the values of $\Lambda$ and $\mu$ for this cosmological application we obtain,
\begin{equation}
  \tau_s \sim 4\, \mathrm{Gyrs} \, ,
\end{equation}
after restoring units.
Meaning that the angular distribution of the angular momenta is already oriented towar the galactic pole after 1 Gyr (as can be seen in Fig.~\ref{fig:orbital_poles}) and that the charge-swapping configurations live for much longer than is required for the galactic halos in the Universe.

\begin{figure*}
	\begin{centering}
	  \includegraphics[width=0.4\textwidth]{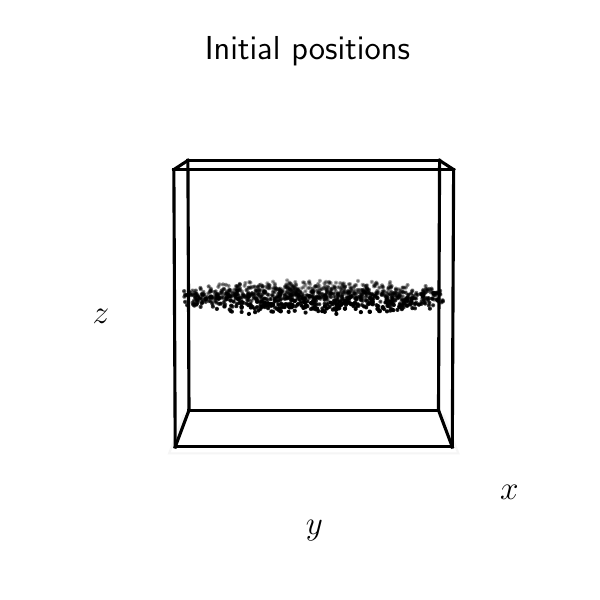}
    \includegraphics[width=0.4\textwidth]{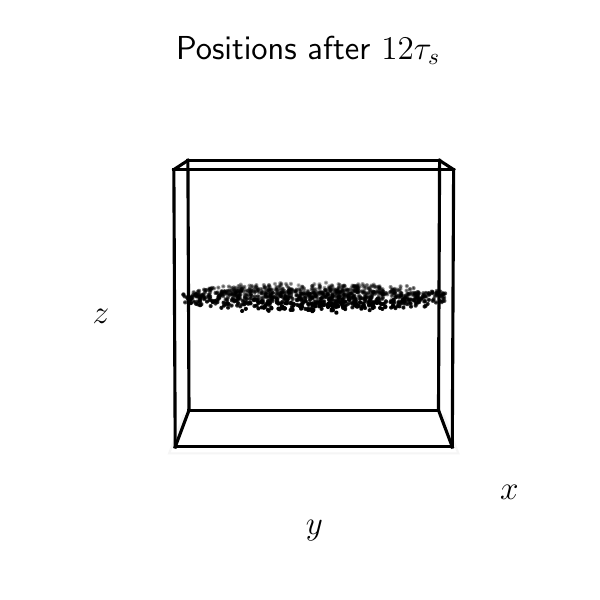}\\
    \includegraphics[width=0.4\textwidth]{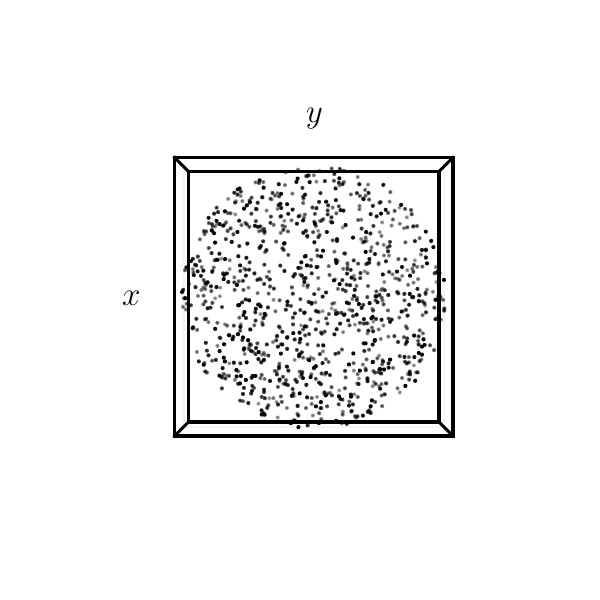}
    \includegraphics[width=0.4\textwidth]{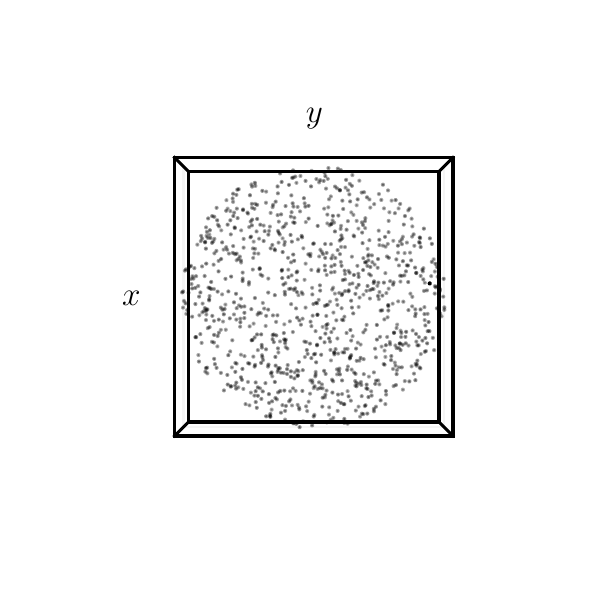}
	\end{centering}
	\caption{``Particles'' in a disc of radius $30$ kpc and thickness $3$ kpc
          moving in a (Newtonian) gravitational potential of the fiducial dipolar configuration (see Fig.~\ref{fig:particle_positions}). The side of the cube is of length $60$ kpc.
          The bottom panels display a top-view of the disc.
        }
	\label{fig:discs}
\end{figure*}

Furthermore, it is possible that the gravitational potential in this scenario could
destroy the Galactic disc. To see that this does not happen, we place ``particles'' in a disk of radius 30 kpc
with a small thickness varying between 0 and 10 percent of the disk radius.
We give initial velocities to the ``particles'' with directions parallel to the
equatorial plane and with velocities such that they would follow perfect circular motion
if they were located in their positions projected in the equatorial plane.
Then, we let them evolve for $12\tau_s$.
We find that within the chosen parameters the structure of the disc does
not change: its radius and thickness remain the same, as can be seen in Fig.~\ref{fig:discs}

\section{Conclusions}
\label{sec:conclusions}

In this paper, we have discussed the complex structures of boson stars and established the existence of charge-swapping configurations in the presence of gravity. We have shown how the metric field changes the stability and other properties of the charge-swapping configurations and recovered the Minkowski flat results in suitable limits. Particularly, the quadratic monomial potential, the sextic polynomial potential and the running mass/logarithmic potential are explored with fully nonlinear numerical relativity simulations. For the quadratic and logarithmic case, the complex boson stars are found to be very stable, and we have yet to see them decay in our long-term simulations. The existence of charge-swapping configurations for the quadratic potential is a novelty when coupling the U(1) scalar to gravity, thanks to  the gravitational attractions. 

In the free scalar case, taking the dipolar boson star for example, we have found that the real and imaginary parts of the scalar field oscillate with slightly different frequencies. This difference determines the main frequency at which the Noether charge of the two components of the system is exchanged. Also, we have found that configurations can avoid gravitational collapse when the initial stars are prepared with sufficiently low masses and separations, as any excess energy is radiated in the initial relaxation stage. In the sextic potential case, we have found evidence of transient existence of compact self-gravitating charge-swapping configurations. Although defining compactness in the case of dynamical configurations is not straightforward, we found that charge-swapping configurations prepared with boson stars with $C \sim 0.1$, some transient configurations emerge after losing about half of their mass (Fig.~\ref{fig:results_E2}). These resulting structures maintain roughly the same radius as the initial stars, rendering the newly formed configurations compact. Thus, a low scalar field density is not a requirement for the existence of these configurations. 

Finally, we have proposed a concrete application for these charge-swapping configurations at large/galactic scales. Assuming that the Galactic dark matter halo is a charges-swapping dipole, we have shown that test particles placed on it will cluster anisotropically and particularly their angular momenta will be oriented in the direction of the Galactic plane. This is different from the scenario of the cold dark matter model where an isotropic halo is more likely to be formed.
We have chosen the model parameters to preliminarily fit the observations in the Milky Way, but a precise determination of the model parameters requires a statistical consistency analysis, which is beyond the scope of the current paper. 

A particular extension of the current work is to study the gravitational signals associated with the relaxation period, after, say, a collision, and the decay or collapse of the most compact charge-swapping configurations, which is also left for future work.

\begin{acknowledgments}
We would like to thank Qi-Xin Xie for helpful discussions. SYZ acknowledges support from the National Key R\&D Program of China under grant No.~2022YFC2204603 and from the National Natural Science Foundation of China under grant No.~12075233 and 12247103.

\end{acknowledgments}

\appendix
\setcounter{secnumdepth}{0}
\section{Appendix: Convergence tests}\label{sec:tests}

The charge-swapping configurations found in this work
are very long-lived, so convergence tests of the numerical code
are very important. First, we have checked the consistency
of the full implementation by constructing and evolving
isolated stable boson stars that are previously known, calculating the
frequency of the scalar field during the simulation and
checking whether it coincides with the frequency of the stationary
solution for a long period of time, $t\sim 10^4$.
The coordinates chosen are such that the different
metric functions are expected to stay static during the evolution,
which we again verify by plotting the global minima and maxima of
some of them. To evaluate the consistency of the charge-swapping
dynamics, we have compared with the
flat spacetime results in \cite{Xie:2021glp} for the polynomial potential
case and \cite{Hou:2022jcd} for the logarithmic potential. Specifically,
we compared the curves of $E$ and found good agreement
despite the fact that the implementation is completely
different.

\begin{figure}
  \centering
	  \includegraphics[width=0.45\textwidth]{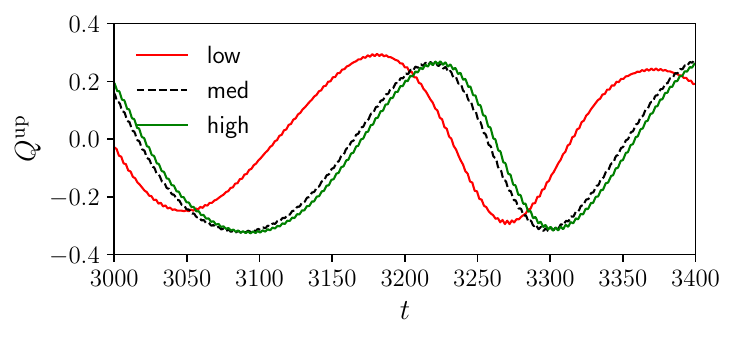}\\
          \includegraphics[width=0.45\textwidth]{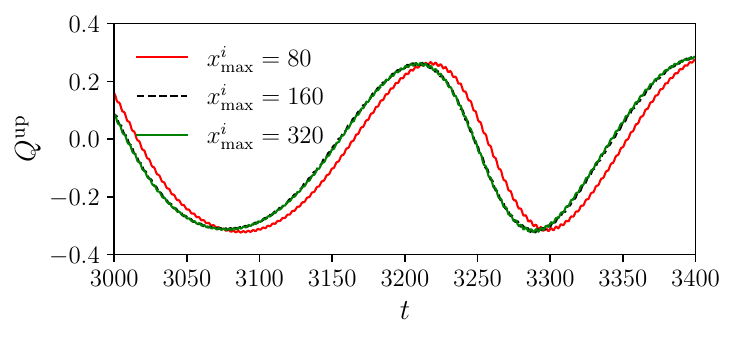}
	\caption{Resolution comparison for the charge $Q^{\rm up}$ for resolutions
          in the finest refinement grid corresponding to
          $\Delta x^i = 0.8$, 0.4 and 0.2 (top panel)
          and different physical box sizes (bottom). For the resolution comparison,
          we use a box size of $160^3$, while for the size comparison we use the medium
          resolution. The configuration corresponds to the free field dipole with
          $\omega = 0.97$ and $d=24$.
        }
	\label{fig:convergenceQ}
\end{figure}

\begin{figure}
	\centering
	  \includegraphics[width=0.45\textwidth]{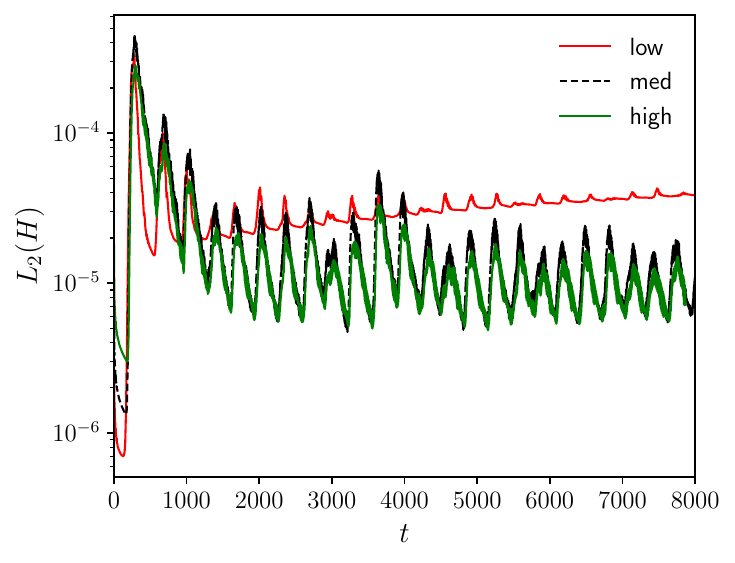}
	\caption{Convergence behavior of the Hamiltonian constraint violation
          for the same case described in Fig.~\ref{fig:convergenceQ}.
        }
	\label{fig:convergenceH}
\end{figure}

In Fig.~\ref{fig:convergenceQ}, we show the Noether charge for
different sizes of the physical box and spatial resolutions.
In this plot we have plotted the time window between $t=3000$ and $t=3400$ to make convergence readily apparent, however the following conclusions can be obtained from other time windows.
We see that the medium and high resolutions
provide precise results. The same applies to the two bigger box cases,
with $x_{\max}^i=160$ and 320, but the three box sizes are in good
agreement for the amplitudes, and the frequencies of the charge
oscillations differ from each other only by a phase.
In our simulations of the dipolar free field case, we used
the medium resolution and the smaller box size. In Fig.~\ref{fig:convergenceH}
we show the $L_2$-norm of the Hamiltonian constraint Eq.~\ref{eq:Hamiltonian}.
The violation of this constraint begins from small values and increases in the initial relaxation stage. The errors
reduce by one order of magnitude for the medium and high resolutions.
If we start with different values of $d$ or start with different models of initial boson stars the violation of the constrains (at least in the medium resolution) is of the same order as the one showed in Fig.~\ref{fig:convergenceH} whenever the configuration do not collapse. For instance, for smaller separations, the constraint violations remain of the same order within a certain range of closely spaced configurations, provided they do not collapse. However, in configurations with initially larger separations, the quantities $H$ and $\mathcal{M}_i$ become more pronounced during the initial relaxation stage.

Finally we comment on the momentum constraint. We have tracked the evolution of it and found that the components $L_2(\mathcal{M}_x)$ and $L_2(\mathcal{M}_y)$ exhibit the same qualitative behavior as the Hamiltonian evolution shown in Fig.~\ref{fig:convergenceH}. The maximum occurs at the beggining of the initial relaxation process, followed by some dissipation of the numerical error. Both components are an order of magnitude smaller than the Hamiltonian and the $L_2(\mathcal{M}_z)$ component is approximately half the size of the $x$ and $y$ components.

\bibliography{ref}

\end{document}